\documentclass[journal=jpcafh,manuscript=article]{achemso}

\usepackage[version=3]{mhchem} 

\author{Aleksander P. Woźniak}
\email{ap.wozniak@uw.edu.pl}
\author{Robert Moszyński}
\affiliation[Unknown University]
{Faculty of Chemistry, University of Warsaw, Pasteura 1, 02-093 Warsaw, Poland}

\title[An \textsf{achemso} demo]
  {Modeling of high harmonic generation in the C\textsubscript{60} fullerene using \textit{ab initio}, DFT-based and semiempirical methods }

\abbreviations{IR,NMR,UV}

\begin{document}

\begin{tocentry}

\includegraphics[width=1.\linewidth]{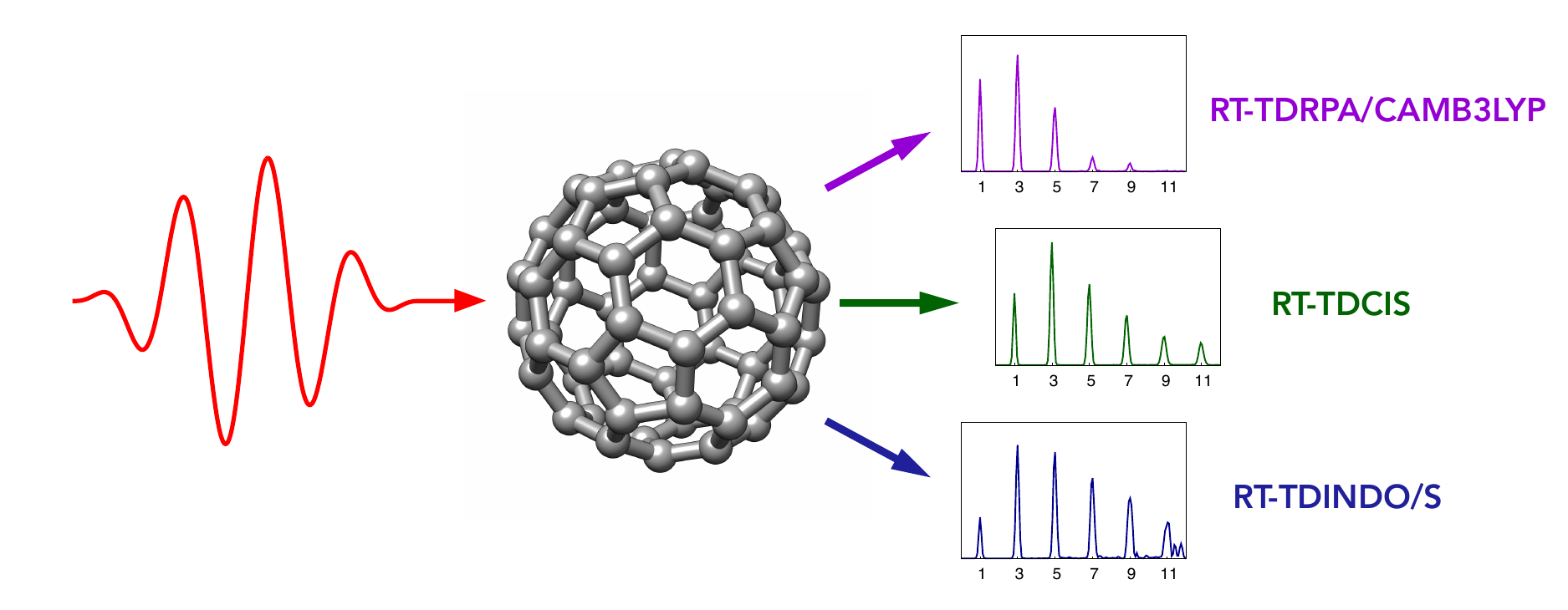}

\end{tocentry}

\begin{abstract}

We report calculations of the high harmonic generation spectra of the C\textsubscript{60} fullerene molecule, employing a diverse set of real-time time-dependent quantum chemical methods.
All methodologies involve expanding the propagated electronic wavefunction in bases consisting of the ground and singly excited time-independent eigenstates obtained through the solution of the corresponding linear-response equations.
We identify the correlation and exchange effect in the spectra by comparing the results from methods relying on the Hartree-Fock reference determinant with those obtained using approaches based on the density functional theory with different exchange-correlation functionals.
The effect of the full random phase approximation treatment of the excited electronic states is also analyzed and compared with the configuration interaction singles and the Tamm-Dancoff approximation.
We also showcase that the real-time extension of the semiempirical method INDO/S can be effectively applied for an approximate description of laser-driven dynamics in large systems.

\end{abstract}

\section{Introduction}

When atoms and molecules are subjected to extremely intense laser fields, their electron densities undergo rapid oscillations, emitting electromagnetic radiation that contains up to multiple hundreds of harmonic frequencies of the incident light.
The discovery of this phenomenon, known as high-harmonic generation (HHG), has ushered in the era of attophysics \cite{agostini2004,krausz2009}.
It has enabled the routine production of ultrashort coherent electromagnetic pulses, making it possible to explore electron dynamics on previously unattainable timescales.
Attosecond impulses have a wide range of applications, including, among others, molecular imaging \cite{vozzi2011,salieres2012,peng2019}, monitoring chemical and photochemical reactions in real time \cite{worner2010a,baykusheva2019}, determining molecular structures \cite{torres2010,wong2011}, studying photoionization \cite{worner2009,bruner2016}, and investigating quantum coherence and electronic correlation effects \cite{worner2010b,shiner2011,marciniak2019,monfared2022}.

HHG was initially observed in noble \cite{mcpherson1987,ferray1988,li1989,sarukura1991,crane1992,faldon1992,miyazaki1992,lhuillier1993,kondo1993,macklin1993,wahlstrom1993} and molecular \cite{liang1994,chin1995,lynga1996} gases, with the former remaining the most commonly used sources of harmonic radiation.
In simple mono- or few-atomic systems a single HHG event can be explained with the well-known three-step model (3SM) \cite{krause1992a,krause1992b,schafer1993,corkum1993,lewenstein1994}, according to which an electron (1) is detached from the atom or molecule via tunneling ionization, (2) is accelerated away by the driving field and then reaccelerated back toward the parent ion when the field switches its sign and (3) recombines with the parent ion which leads to the emission of radiation burst.
High-harmonic generation from gas targets, although relatively easy to achieve, is however hindered by a low conversion efficiency attributed to the low density of the medium \cite{constant1999}.
Thus, there is an ongoing search for novel, more efficient HHG sources.
In recent years, thanks to the rapid development of experimental techniques, HHG was demonstrated to occur also in bulk solids \cite{ghimire2011,schubert2014,luu2015,ndabashimiye2016,you2017}, liquids \cite{luu2018}, and nanostructures \cite{golde2008,singhal2010,han2016,liu2017,sivis2017,shcherbakov2021}, with the harmonic yield greatly exceeding that of atomic and molecular gases.
The mechanism of harmonic generation in bulk media, although not fully revealed, is suspected to differ substantially from the 3SM \cite{ghimire2011,luu2015}.

Progress in experimental discoveries in attophysics also necessitates the development of new theoretical methods capable of describing ultrafast electron dynamics in increasingly larger systems.
In the past decade, there has been a significant rise in the popularity of employing quantum chemistry methods extended to the real-time domain for this purpose \cite{lopata2011,ishikawa2015,goings2017,bedurke2021,coccia2022}.
These approaches are characterized by moderate computational costs typical of their time-independent counterparts, as well as a reasonable level of accuracy, allowing, for example, the consideration of multi-electron effects.
One of the most popular methods of this kind is the real-time time-dependent configuration interaction singles (RT-TDCIS), in which the time-dependent electronic wavefunction is represented as a linear combination of time-independent ground and singly excited electronic eigenstates of the examined system \cite{klamroth2003,krause2005,huber2005,rohringer2006,krause2007,schlegel2007}.
Thanks to its highly favorable scaling with the number of electrons, it can be routinely applied not only to atoms \cite{luppi2013,coccia2016a,coccia2016b,pabst2016,coccia2019,coccia2020,wozniak2021,wozniak2022} and simple molecules \cite{luppi2012,white2016,labeye2018,pauletti2021,wozniak2023}, but also to complex organic \cite{bedurke2019} and biological \cite{luppi2021,morassut2022,luppi2023} compounds, often yielding results qualitatively or even quantitatively consistent with the experimental data \cite{luppi2021,bedurke2019,wozniak2023b}.
This raises the question of whether it can perform equally well for even larger systems, such as nanostructured materials.

In this work we report calculations of the HHG spectra of arguably the most well-known carbon nanostructure, the C\textsubscript{60} fullerene molecule, using quantum chemical approaches coupled to Gaussian basis sets.
Fullerenes are currently of great interest of attophysics, as experimental studies report their exceptionally high HHG yield \cite{ganeev2009a,ganeev2009b,ganeev2009c,ganeev2013,ganeev2016}, significantly surpassing that of bulk carbon \cite{ganeev2009a,ganeev2009c,ganeev2016}.
This property is attributed to their high polarizability, as well as the  occurence of the plasmon resonance at the fullerene surface \cite{ganeev2011}.
From the theoretical point of view, HHG in fullerenes has been studied either using an extention of the three-step model \cite{ciappina2007,ciappina2008}, or by real-time simulations employing a variety of different approximate Hamiltonians.
These include SAE-based models \cite{ganeev2013,topcu2019}, tight-binding models \cite{zhang2005,zhang2006,avetissian2021}, jelliumlike approximation \cite{redkin2010} and density functional theory combined with pseudopotentials \cite{redkin2011,zhang2020}.
However, for most of these models to perform properly, some form of system-specific parametrization is typically necessary, such as the construction of effective potentials.
On the other hand, all-electron quantum chemical methods are much less system-dependent and offer a more universal simulation framework, which may prove very useful in possible future studies, such as investigating the role of various chemical modifications on HHG in nanostructures.
Therefore, the first objective of this work is to assess whether the applicability of RT-TDCIS can be extended to large structures containing tens of atoms.
RT-TDCIS, being an equivalent of the Hartree-Fock (HF) method for excited states, does not account for correlation effects. 
Although it has been demonstrated that dynamical electron correlation has little effect on the laser-driven dynamics in atoms \cite{castro2015,artemyev2017,sato2018,neufeld2020,reiff2020,wozniak2022} and small molecules \cite{nguyen2006,neufeld2020,saalfrank2020,pauletti2021,wozniak2023}, in the case of C\textsubscript{60} it is known that the single-determinant restricted HF (RHF) wavefunction is not a stable ground state due to global correlations in the $\pi$ orbital space \cite{stuck2011}.
Therefore, limiting the calculations solely to RT-TDCIS may not be sufficient to obtain reliable HHG spectra.
Among several quantum chemical approaches that incorporate dynamical correlation into the real-time dynamics, the most popular is the real-time time-dependent density functional theory (RT-TDDFT), in which the propagation is applied directly to Kohn-Sham spinorbitals forming the electron density \cite{runge1984,tong1998,castro2004}.
While including correlation effects implicitly through the use of the exchange-correlation potential of choice, RT-TDDFT suffers from its own limitations, such as an inability to describe single excitations in closed-shell systems \cite{isborn2008,habenicht2014}, non-linearity of the time-evolution equations \cite{castro2012}, and, last but not least, computational costs greatly exceeding those of RT-TDCIS \cite{bedurke2021}.
The latter drawback is also shared with more sophisticated multideterminant methods such as real-time time-dependent coupled cluster (RT-TDCC) \cite{kvaal2012}, RT-TDCISD \cite{krause2007,schlegel2007,saalfrank2020,wozniak2022,wozniak2023} and RT-TDCIS(D)\cite{krause2005,krause2007,schlegel2007}.
An alternative approach to RT-TDDFT, proposed by Pauletti \textit{et al.} \cite{pauletti2021} and also utilized in this work, is to add the exchange-correlation potential directly to the RT-TDCIS Hamiltonian, effectively turning it into the real-time time-dependent counterpart of the Tamm-Dancoff approximation (TDA).
Therefore, we compute the HHG spectra of the C\textsubscript{60} molecule using RT-TDCIS and RT-TDTDA, as well as their respective generalizations that include deexcitation terms: the real-time time-dependent extensions of the linear-response time-dependent Hartee-Fock method and the linear-response time-dependent density functional theory.
This allows us to assess the role of multi-electron effects in the laser-driven dynamics in fullerenes, which constitutes the second goal of this paper.
Finally, we also calculate the HHG response using the semiempirical INDO/S Hamiltonian, which has been parameterized to reproduce excitation energies obtained with CIS in a limited active space \cite{ridley1973}.
INDO/S has been recently extended to the real-time domain, but so far, it has only been employed in the modeling of absorption spectra of in the perturbative regime, giving somewhat promising results \cite{ghosh2017}.
Thus, the third and final aim of this paper is to investigate if INDO/S can also be applied for simulating strong field dynamics in large systems and serve as a less expensive alternative to \textit{ab initio} and DFT-based methods.

The paper is constructed as follows.
In Sec. II we present a brief theoretical background of the used methods, as well as provide computational details of the simulations.
In Sec. III we present and discuss the results of the HHG calculations on the C\textsubscript{60} fullerene.
Finally, in Sec. IV we conclude our work.

\section{Methods}

\subsection{Theory}

In this section we provide an overview of the theoretical foundations for the methods used in the present work, collectively referred to as the real-time time-dependent single excitation-based methods (RT-TDSEM).
The common feature among all RT-TDSEM is providing an approximate solution to the TDSE,
\begin{equation} \label{tdse}
i \frac{\partial}{\partial t}\Psi(t) = \hat{H}(t)\Psi(t),
\end{equation}
by expanding the time-dependent electronic wavefunction in the basis of the time-independent eigenstates $\Psi_{m}$ of the examined system, which include the ground state $\Psi_0$ and the excited states $\Psi_{k}$,
\begin{equation} \label{tdwf}
\Psi(t) = \sum_{m} C_m(t)\Psi_m = C_0(t)\Psi_0 + \sum_{k>0} C_k(t) \Psi_k.
\end{equation}
In all considered RT-TDSEM $\Psi_0$ is assumed to be the ground state closed-shell Slater determinant $\Phi_0$ built from real, orthonormal occupied molecular orbitals (MOs) $\phi_i$ represented in the linear combination of atomic orbitals (LCAO) approximation,
\begin{equation} \label{lcao}
\phi_i = \sum_\mu c_{\mu i} \chi_\mu
\end{equation}
In most electronic structure methods, Gaussian functions are chosen for the atomic orbital basis set $\chi_\mu$ due to computational efficiency.
The molecular orbitals $\phi_i$ are the solutions to the Hartree-Fock or Kohn-Sham self-consistent field (SCF) equations,
\begin{equation} \label{scf}
\mathbf{Fc} = \mathbf{ScE}
\end{equation}
where $\mathbf{F}$ is either the Fock or Kohn-Sham matrix, $\mathbf{S}$ is the overlap matrix, $\mathbf{c}$ is the matrix of coefficients $c_{\mu i}$, and $\mathbf{E}$ is the diagonal matrix of MO energies.
In addition to the occupied MOs, solving the SCF problem also provides the set of virtual MOs $\phi_a$.

The excited states $\Psi_k$ in Eq. \eqref{tdwf} are obtained by solving a linear-response equation specific to the particular RT-TDSEM.
The arguably most general linear-response theory considered in this work is the linear-response time-dependent density functional theory (LR-TDDFT), based on the Kohn-Sham reference determinant.
The (non-Hermitian) LR-TDDFT eigenequation reads \cite{dreuw2005}
\begin{equation} \label{lrtddft}
\begin{pmatrix}
\mathbf{A} & \mathbf{B}  \\
-\mathbf{B} & -\mathbf{A}  \\
\end{pmatrix}
\begin{pmatrix}
\mathbf{X} \\
\mathbf{Y} \\
\end{pmatrix}
 = 
\omega
\begin{pmatrix}
\mathbf{X} \\
\mathbf{Y} \\
\end{pmatrix}
\end{equation}
Here, the matrices $\mathbf{A}$ and $\mathbf{B}$ are commonly referred to as the excitation and deexcitation matrix, respectively.
Their elements are defined as
\begin{gather} 
A_{ia,jb} = \delta_{ij}\delta_{ab}(\epsilon_a - \epsilon_i) + (ia|jb) + (ia|f_{xc}|jb) \\ 
B_{ia,jb} = (ia|bj) + (ia|f_{xc}|bj)
\end{gather}
where $i$ and $j$ denote occupied MOs (hole states) $\phi_i$ and $\phi_j$, $a$ and $b$ denote virtual MOs (particle states) $\phi_a$ and $\phi_b$, $\epsilon_i$ and $\epsilon_a$ are the occupied and virtual MO energies, respectively, and $f_{xc}$ is the exchange-correlation kernel used in the SCF procedure.
The two-electron Coulomb integrals $(ia|bj)$ and exchange-correlation integrals $(ia|f_{xc}|bj)$ are written in the Mulliken notation.
According to Casida \cite{casida1995}, every excited state $\Psi_k$ with the excitation energy $\omega_k = E_k - E_0$ is defined by the vectors $\mathbf{X}_k$ and $\mathbf{Y}_k$ as
\begin{equation} \label{state}
\Psi_k = \sum_i^{occ} \sum_a^{vir} (X_{ia}^k + Y_{ia}^k) \Phi_i^a,
\end{equation}
where $\Phi_i^a$ constitute the set of singly excited Slater determinants.
The vectors $\mathbf{X}_\nu$ and $\mathbf{Y}_\nu$ also fulfill the normalization condition
\begin{equation} \label{norm}
\sum_i^{occ} \sum_a^{vir} (X_{ia}^k X_{ia}^l - Y_{ia}^k Y_{ia}^l) = \delta_{kl}.
\end{equation}

If the orbitals used to construct the matrices $\mathbf{A}$ and $\mathbf{B}$ are obtained from the Hartree-Fock method, the exchange-correlation kernel $f_{xc}$ is replaced by the (non-local) HF exchange kernel $f_\mathrm{HF}$, $(ia|f_{xc}|bj) = (ia|f_\mathrm{HF}|bj) = -(ib|aj)$, and Eq. \eqref{lrtddft} is reduced to the linear-response time-dependent Hartree-Fock (LR-TDHF) method, also known as the random phase approximation (RPA) \cite{dreuw2005}.
It is worth mentioning that despite their similarities, LR-TDHF and LR-TDDFT have, in fact, been derived independently from each other, with the former being significantly older \cite{mclachlan1964,rowe1968}.
However, in this work it is more convenient for us to treat LR-TDHF as a special case of LR-TDDFT, since we aim at investigating the effect of the exchange-correlation potential on the laser-driven electron dynamics.

The terms LR-TDHF and LR-TDDFT have traditionally been used for the methods solely for obtaining the excited eigenspectra of atoms and molecules.
On the other hand, RT-TDHF and RT-TDDFT usually refer to the single-determinant approaches in which the real-time propagation is applied to molecular orbitals.
Moreover, some authors use the term RPA interchangeably not only with LR-TDHF, but also with LR-TDDFT \cite{fabiano2012,ziegler2014,hesselmann2015}.
In order to avoid possible confusion, we make use of this fact in this work and refer to the method of propagating the time-dependent wavefunction \eqref{tdwf}, expanded in the basis of states obtained through the solution of Eq. \eqref{lrtddft}, as RT-TDRPA/xc, where xc may stand for HF, DFT, or any particular exchange-correlation functional.

Setting the deexcitation matrix $\mathbf{B}$ to zero in the LR-TDDFT (LR-TDHF) equation leads to the well-known TDA (CIS) approximation \cite{dreuw2005},
\begin{equation} \label{cis}
\mathbf{A}\mathbf{X} = \omega\mathbf{X},
\end{equation}
The properties \eqref{state} and \eqref{norm} also apply to TDA and CIS states, with the exception that $\mathbf{Y}_k = 0$ for every $\Psi_k$.
TDA (CIS) approximation already provides a significant simplification of the linear-response problem compared to the full LR-TDDFT (LR-TDHF), as Eq. \eqref{cis} is Hermitian.
This allows for a reduction in computational costs required to obtain excited states, as well as helps avoid various numerical issues, with the most infamous being the triplet instabilities \cite{cizek1967}.
However, TDA (CIS) still necessitates calculations of all the one- and two-electron integrals required to solve the ground state Kohn-Sham (Hartree-Fock) equations and to construct the excitation matrix $\mathbf{A}$, which can be prohibitive for very large systems.

To overcome this bottleneck, various semiempirical quantum chemical methods have been historically proposed, all of which rely on setting certain types of integrals to zero or replacing them with much simpler parameterized formulae.
Although the parameters of most semiempirical methods have been adjusted to reproduce the results of \textit{ab initio} ground state calculations \cite{thiel2014}, one of them, named INDO/S, has been specifically constructed to yield excitation energies matching those of CIS in a truncated active orbital space \cite{ridley1973,voityuk2013}.
The detailed description of INDO/S can be found elsewhere \cite{ridley1973,ridley1976,bacon1979,zerner1980}, while here we briefly discuss only the key assumptions of this method.
Unlike in the \textit{ab initio} and DFT-based methods, and similarly to other semiempirical methods, the MOs in INDO/S are expanded not in a Gaussian basis set, but in the minimal basis set of Slater orbitals that describe only the valence orbitals of every element.
INDO/S originates from the Hartree-Fock formalism, with the elements of the Fock matrix defined in the atomic orbital representation as
\begin{equation} \label{fock}
F_{\mu\nu} = h_{\mu\nu} + \sum_{\lambda\sigma} P_{\lambda\sigma} \left( (\mu\nu|\lambda\sigma) - \frac{1}{2}(\mu\nu|\lambda\sigma) \right),
\end{equation}
where $h_{\mu\nu}$ are the elements of the one-electron Hamiltonian matrix and $P_{\lambda\sigma}$ are the elements of the one-electron density matrix, $P_{\lambda\sigma} = 2\sum_i^{occ} c_{\lambda i} c_{\sigma i}$
However, the solution of the SCF problem is heavily simplified due to the so-called zero differential overlap (ZDO) approximation, which sets the overlap matrix $\mathbf{S}$ in Eq. \eqref{scf} to an identity matrix, $S_{\mu\nu} = \delta_{\mu\nu}$.
As a consequence, all three- and four-center two-electron integrals between basis functions $(\mu\nu|\lambda\sigma)$ vanish.
The two-center two-electron are also set equal to zero with the exception of Coulomb-like integrals involving only two basis functions, $(\mu^A\nu^A|\lambda^B\sigma^B) = \delta_{\mu^A\nu^A}\delta_{\lambda^B\sigma^B}(\mu^A\mu^A|\lambda^B\lambda^B)$, which are calculated from the Mataga-Nishimoto formula \cite{mataga1957} (the superscripts $A$ and $B$ denote different atomic centers).
The one-center two-electron integrals are estimated using the Slater-Condon Coulomb and exchange factors.
The one-electron integrals $h_{\mu\nu}$ are also approximated using combinations of one-electron core integrals, resonance integrals, modified overlap integrals, and two-center two-electron integrals \cite{ridley1973}.
All approximations used during the construction of the semiempirical Fock matrix are also applied when constructing the INDO/S excitation matrix $\mathbf{A}$.

Having obtained the set of electronic states $\Psi_k$ using the linear-response theory of choice, one can proceed to propagate the time-dependent wavefunction $\Psi(t)$.
The time-dependent Hamiltonian $\hat{H}(t)$ in Eq. \eqref{tdse} consists of the ground state molecular Hamiltonian $\hat{H}_0$ and the interaction operator coupling the molecule to the external laser field, represented in the length gauge and within the dipole approximation,
\begin{equation} \label{hamiltonian}
\hat{H}(t) = \hat{H}_0 - \vec{\hat{\mu}}\vec{\mathcal{E}}(t),
\end{equation}
where $\vec{\hat{\mu}}$ is the molecular dipole operator and $\vec{\mathcal{E}}(t)$ is the time-dependent external electric field.
Inserting Eqs. \eqref{tdwf} and \eqref{hamiltonian} into Eq. \eqref{tdse} leads to the equation for the time-evolution of the state vector $\mathbf{C}(t)$ of time-dependent coefficients $C_k(t)$,
\begin{equation} \label{time-evolution}
i \frac{\partial}{\partial t}\mathbf{C}(t) = \left(\mathbf{H}_0 + \sum_{\alpha = x,y,z}  \mathcal{E}_\alpha(t) \boldsymbol{\mu}_\alpha \right) \mathbf{C}(t).
\end{equation}
Here, $\mathbf{H}_{0,mn} = \delta_{mn} E_m$ and $\boldsymbol{\mu}_\alpha$ is the matrix of the $\alpha$-th spatial component of the dipole moment operator, $\boldsymbol{\mu}_{\alpha,mn} = \langle \Psi_m | \mu_\alpha \Psi_n \rangle$.
The elements of $\boldsymbol{\mu}_\alpha$, in particular the dipole moment expectation values of excited states $\boldsymbol{\mu}_{\alpha,kk}$ and the dipole transition moments between two excited states $\boldsymbol{\mu}_{\alpha,kl}$, can be determined with high accuracy using quadratic response theory \cite{cronstrand2000}.
However, since we aim at propagating the time-dependent wavefunction using the full eigenspectrum of Eq. \eqref{lrtddft}, which for systems as large as C\textsubscript{60} may consist of tens of thousands of states, this approach would be practically unfeasible.
Therefore, we calculate the elements of $\boldsymbol{\mu}_\alpha$ in an approximate manner, by inserting Eq. \eqref{state} directly into $\langle \Psi_m | \mu_\alpha \Psi_n \rangle$ \cite{yeager1975}:
\begin{gather}
\langle \Psi_0 | \mu_\alpha \Psi_k \rangle = \sum_i^{occ} \sum_a^{vir} \left( X_{ia}^k + Y_{ia}^k \right) \langle \Phi_0 | \mu_\alpha \Phi_i^a \rangle,  \label{gerpa} \\
\langle \Psi_k | \mu_\alpha \Psi_l \rangle = \sum_{ij}^{occ} \sum_{ab}^{vir} \left( X_{ia}^k X_{jb}^l + Y_{ia}^k Y_{jb}^l \right) \langle \Phi_i^a | \mu_\alpha \Phi_j^b.  \label{eerpa} \rangle
\end{gather}
The dipole moment integrals between the ground and excited Slater determinants can be readily evaluated using Slater-Condon rules.
For CIS, TDA, and INDO/S states these expressions reduce to
\begin{gather}
\langle \Psi_0 | \mu_\alpha \Psi_k \rangle = \sum_i^{occ} \sum_a^{vir} X_{ia}^k \langle \Phi_0 | \mu_\alpha \Phi_i^a \rangle, \label{gecis} \\
\langle \Psi_k | \mu_\alpha \Psi_l \rangle = \sum_{ij}^{occ} \sum_{ab}^{vir} X_{ia}^k X_{jb}^l \langle \Phi_i^a | \mu_\alpha \Phi_j^b. \label{eecis} \rangle
\end{gather}

To solve Eq. \eqref{time-evolution} we introduce time discretization and propagate the wavefunction using the second-order split-operator technique,
\begin{equation} \label{propagator}
\mathbf{C}(t+\Delta t) = e^{i \mathbf{H_0}\Delta t/2} \left[ \prod_{\alpha=x,y,z} \mathbf{U}_\alpha e^{i \mathcal{E}_\alpha(t)\mathbf{d}_\alpha \Delta t} \mathbf{U}_\alpha^\dag \right] e^{i \mathbf{H_0}\Delta t/2} \mathbf{C}(t),
\end{equation}
where $\Delta t$ denotes the time step and the unitary matrix $\mathbf{U}_\alpha$ diagonalizes the $\alpha$-th dipole component matrix, $\mathbf{U}_\alpha \boldsymbol{\mu}_\alpha \mathbf{U}_\alpha^\dag = \mathbf{d}_\alpha$.

Similarly to our previous works \cite{wozniak2021,wozniak2022,wozniak2023,wozniak2023b} and to the works of other authors utilizing RT-TDSEM to simulate strong field electron dynamics \cite{luppi2013,coccia2016a,coccia2016b,labeye2018,bedurke2019,coccia2019,luppi2021,pauletti2021,morassut2022,luppi2023}, we employ the heuristic finite lifetime model of Klinkusch \textit{et al.} \cite{klinkusch2009} to compensate for the incompleteness of the atomic orbital basis sets.
The electronic energies $E_k$ (the diagonal entries of $\mathbf{H}_0$) of excited states beyond the ionization threshold are modified by adding imaginary ionization rates,
\begin{equation}
E_k \rightarrow \overline{E}_k = E_k - i \Gamma_m /2 \qquad \text{for} \; E_k \geq E_0 + I_p.
\end{equation}
The finite lifetime model was originally developed for RT-TDCIS \cite{klinkusch2009}, and later extended to RT-TDCI with higher excitations \cite{coccia2017,wozniak2022}. The ionization rates of CIS states are defined as
\begin{equation} \label{cis_ir}
\Gamma_k = \sum_{i}^{occ.} \sum_{a}^{vir.} |X_{ia}^k|^2 \theta(\epsilon_a) \sqrt{2\epsilon_a}/d,
\end{equation}
where $\theta(x)$ is the Heaviside step function and the empirical parameter $d$ represents a maximum distance from the molecule that a (semiclassical) electron can travel before undergoing ionization.
Naturally, Eq. \eqref{cis_ir} applies to TDA and INDO/S states as well.
In this work, we also extend the finite lifetime model to RT-TDRPA.
By analogy with Eq. \eqref{cis_ir} we define the heuristic ionization rates of LR-TDDFT and LR-TDHF as
\begin{equation} \label{rpa_ir}
\Gamma_k = \sum_{i}^{occ.} \sum_{a}^{vir.} \left(|X_{ia}^k|^2 - |Y_{ia}^k|^2\right) \theta(\epsilon_a) \sqrt{2\epsilon_a}/d.
\end{equation}
The motivation behind Eq. \eqref{rpa_ir} is as follows.
The ionization rate of every CIS state \eqref{cis_ir} can be interpreted as a sum of excitation probabilities to individual virtual orbitals $|X_{ia}^k|^2$ multiplied by the ionization rates of these orbitals $\sqrt{2\epsilon_a}/d$ (the Heaviside function assures that only the virtual MOs with positive energies are ionizable).
Since the RPA theory accounts for both excitations and deexcitations, the ionization rates of LR-TDHF and LR-TDDFT have to be accordingly reduced by the probabilities $|Y_{ia}^k|^2$ that an electron becomes deexcited from the virtual MO $\phi_a$ back to the occupied MO $\phi_i$.

Once the time-dependent wavefunction $\Psi(t)$ is known, the HHG spectrum of the examined system $I_\mathrm{HHG}$ can be calculated from the Fourier transform of the optical response, which in this work is taken to be the dipole acceleration $\ddot{\langle \mu \rangle}$,
\begin{equation} \label{hhg}
I_{\mathrm{HHG}}(\omega) = \left| \frac{1}{T} \int_{0}^{T} \left( \frac{d^2}{dt^2}\sum_{mn} C_m^*(t)C_n(t) \langle \Psi_m|\hat{\mu}\Psi_n \rangle \right) e^{i\omega t}dt \right|^2,
\end{equation}
where $T$ is the total propagation time.

\subsection{Computational details}

We perform the calculations of the HHG spectra of the C\textsubscript{60} molecule subjected to short, intense laser pulses at the RT-TDRPA/DFT, RT-TDRPA/HF, RT-TDTDA, RT-TDCIS, and RT-TDINDO/S levels of theory.
The geometry of C\textsubscript{60} was optimized at the B3LYP-D3/cc-pVTZ level of theory using Gaussian16 (Rev. C.01) software \cite{gaussian}.
The RT-TDRPA/DFT and RT-TDTDA calculations are perfomed in two variants, utilizing two different exchange-correlation functionals.
The first one is the standard B3LYP hybrid functional \cite{becke1993}, frequently employed in calculations involving fullerenes and their derivatives.
The second one is its Coulomb-attenuated version, CAM-B3LYP \cite{yanai2004}, which is reported to perform better than B3LYP in describing excited states \cite{rostov2010,sarkar2021} -- a feature that may be important for the correct description of HHG.
Since we consider the time-evolution of a closed-shell system in the absence of spin-dependent perturbations, we are only interested in the singlet excited manifold, so the linear-response equations are constructed using singlet configuration state functions rather than pure Slater determinants.
We obtain the $\mathbf{A}$ and $\mathbf{B}$ matrices and the dipole moment integrals used in the RT-TDRPA, RT-TDTDA and RT-TDCIS calculations using a modified version of PySCF 2.4 package \cite{pyscf}.
The semiempirical $\mathbf{A}$ matrices used in the RT-TDINDO/S calculations and the dipole moment integrals between INDO/S configurations are generated using a modified version of MOPAC22 \cite{stewart1990,gieseking2021,mopac}.
The solution of the linear-response equation, the construction and diagonalization of the dipole moment matrices $\boldsymbol{\mu}$, and the real-time propagation is performed using a home-made program.
All utilized approaches employ a full diagonalization of either Eq. \eqref{cis} or Eq. \eqref{lrtddft}.
The LR-TDDFT and LR-TDHF equations are solved using the Cholesky decomposition technique \cite{chi1970}.

In the simulations, the external laser field is represented by a linearly polarized electric field pulse with a sine-squared envelope:
\begin{equation} \label{efield}
\vec{\mathcal{E}}(t) = 
\begin{cases}
\vec{\mathcal{E}}_0 \sin(\omega_0 t) \sin^2(\omega_0 t/2 n_c)&\text{if}\; 0 \leq t \leq 2\pi n_c/\omega_0, \\
0& \text{otherwise.}
\end{cases}
\end{equation}
The cycle-averaged laser intensity $I_0$, related to the field amplitude $\mathcal{E}_0$ via $I_0 = \mathcal{E}_0^2/2$, is set to $5\times 10^{13}$ W/cm\textsuperscript{2}, and the number of optical cycles $n_c = 10$.
We use two values of carrier frequency $\omega_0$: 1.55 eV and 0.95 eV, corresponding to the wavelengths of 800 nm and 1300 nm, respectively.
The 800 nm pulse has the duration of 1103 a.u. ($\approx$ 26.7 fs), while the 1300 nm pulse has the duration of 1793 a.u. ($\approx$ 43.4 fs).
Due to the high symmetry of C\textsubscript{60} and computational limitations, we consider only one polarization vector parallel to one of the $S_6$ improper axis.
The wavefunctions are propagated using a timestep $\Delta t = 0.01$ a.u., a value that ensured convergence of obtained results.
Every propagation starts from the single-determinant ground state, which serves as a reference for the corresponding RT-TDSEM.
Also, the value of $I_p$ used in the finite lifetime model is estimated from the Koopmans' theorem, as the negative energy of the highest occupied molecular orbital $-\epsilon_{\mathrm{HOMO}}$ within the reference state corresponding to a particular RT-TDSEM.

The HHG spectra are computed from the dipole acceleration obtained by numerically differentiating the time-resolved dipole moment twice with respect to time.
We also apply the Hann window to the dipole acceleration before taking the Fourier transform in order to account for the finite simulation time.

\section{Results and discussion}

\subsection{Optimization of the HHG simulation framework}

When performing quantum-chemical calculations in the strong-field non-perturbative regime, three factors have the most significant influence on the accuracy of reproducing laser-driven electron dynamics: \\
(a) the employed Gaussian basis set.
Given that multiple transitions to highly excited and unbound electronic states are an inherent part of HHG, it is desired to simulate the electron dynamics using the most accurate available representation of the electronic continuum.
In smaller systems, this typically entails using large numbers of highly diffuse functions, often tailored for this specific purpose \cite{coccia2016a,coccia2016b,wozniak2021,morassut2023}.
It is worth noting that in our case, the choice of the basis set pertains only to the \textit{ab initio} and DFT-based methods, as INDO/S has been designed to work solely with the minimal Slater basis set; \\
(b) the size of the active orbital space.
When the size of the simulated system becomes considerable, a full configurational space with excitations from all occupied MOs to all virtual MOS can no longer be employed in the calculations, especially if a large basis set is used and the propagation involves the entire eigenspectrum of the linear-response equation.
In our earlier calculations on the H\textsubscript{2} molecule we demonstrated that truncating the virtual orbital space may be suboptimal, because excluding the highest lying virtual MOs may negatively affect the description of all excited states, not just those with the highest energies \cite{wozniak2023}.
At the same time, other studies indicate that HHG is usually dominated by transitions from several of the highest lying occupied MOs \cite{luppi2023}, so excluding excitations from core orbitals may be a preferred option for reducing the number of configurations;\\
(c) the parameterization of the applied wavefunction absorber.
In the finite lifetime model employed in our calculations the absorption rate is governed by the escape length value $d$.
Ideally, the absorber should selectively eliminate components of the wavefunction that cannot be accurately represented by the basis set, without interfering with the HHG process.
In practical applications involving atoms and moderately-sized molecules, known to generate harmonics in accordance with the three-step model, a value of $d$ close to the maximum electron excursion distance in the laser field, $E_0/\omega_0^2$, is usually selected.
However, since HHG in nanostructures is known to deviate from the 3SM, the proper choice of $d$ requires additional investigation.

\begin{figure}
  \includegraphics[width=0.5\linewidth]{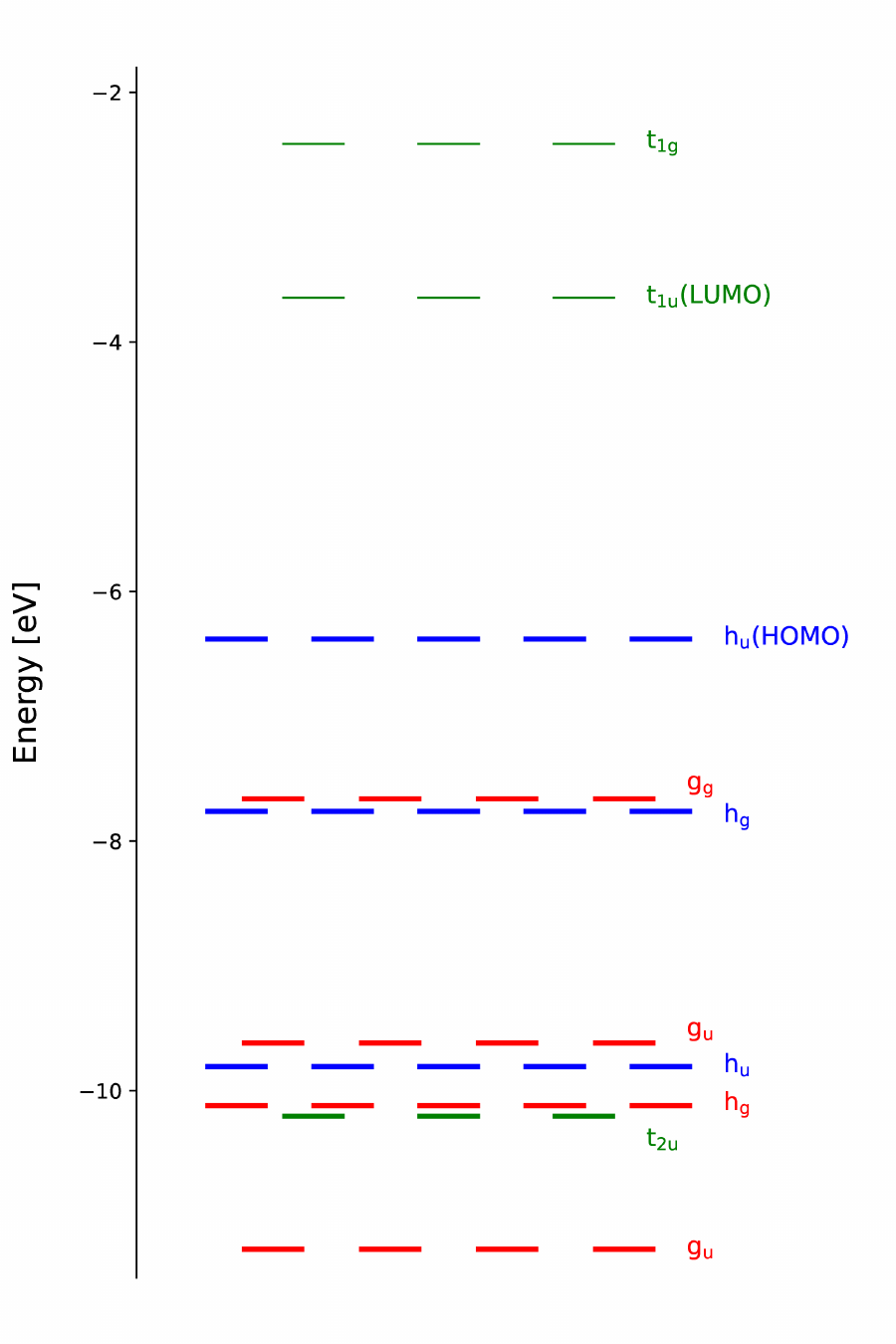}
  \caption{MO energy diagram of 35 highest occupied (thick bars) and 6 lowest unnocupied (thin bars) molecular orbitals of the C\textsubscript{60} molecule, calculated at B3LYP/aug-cc-pVDZ level of theory. The irreps of the $I_h$ symmetry group corresponding to individual orbital groups are also provided}
  \label{fgr:mos}
\end{figure}

Therefore, before comparing the HHG spectra obtained using different RT-TDSEM methods, we must first ensure that the results are converged with respect to all three above parameters.
To achieve this, we conduct a series of benchmark HHG calculations for the C\textsubscript{60} molecule at both considered carrier frequencies, at the RT-TDTDA/B3LYP level of theory.
We test the performance of three basis sets: the minimal STO-3G basis set, the double-zeta Dunning cc-pVDZ basis set, and its singly augmented variant, aug-cc-pVDZ.
Due to the icosahedral symmetry of C\textsubscript{60}, almost all of its MOs belong to degenerate energy levels, with the HOMO level exhibiting fivefold degeneracy (Fig. \ref{fgr:mos}).
We determine the optimal configurational space by employing a full virtual orbital space and gradually expanding the active occupied orbital space, starting from five HOMO orbitals and adding one occupied MO shell at a time.
Finally, we perform every calculation using seven different values of the $d$ parameter: 10, 50, 100, 150, 200, 250 and 300 bohr, and with the finite lifetime model turned off (which is equivalent to setting $d \rightarrow \infty$).

\begin{figure}
  \includegraphics[width=0.99\linewidth]{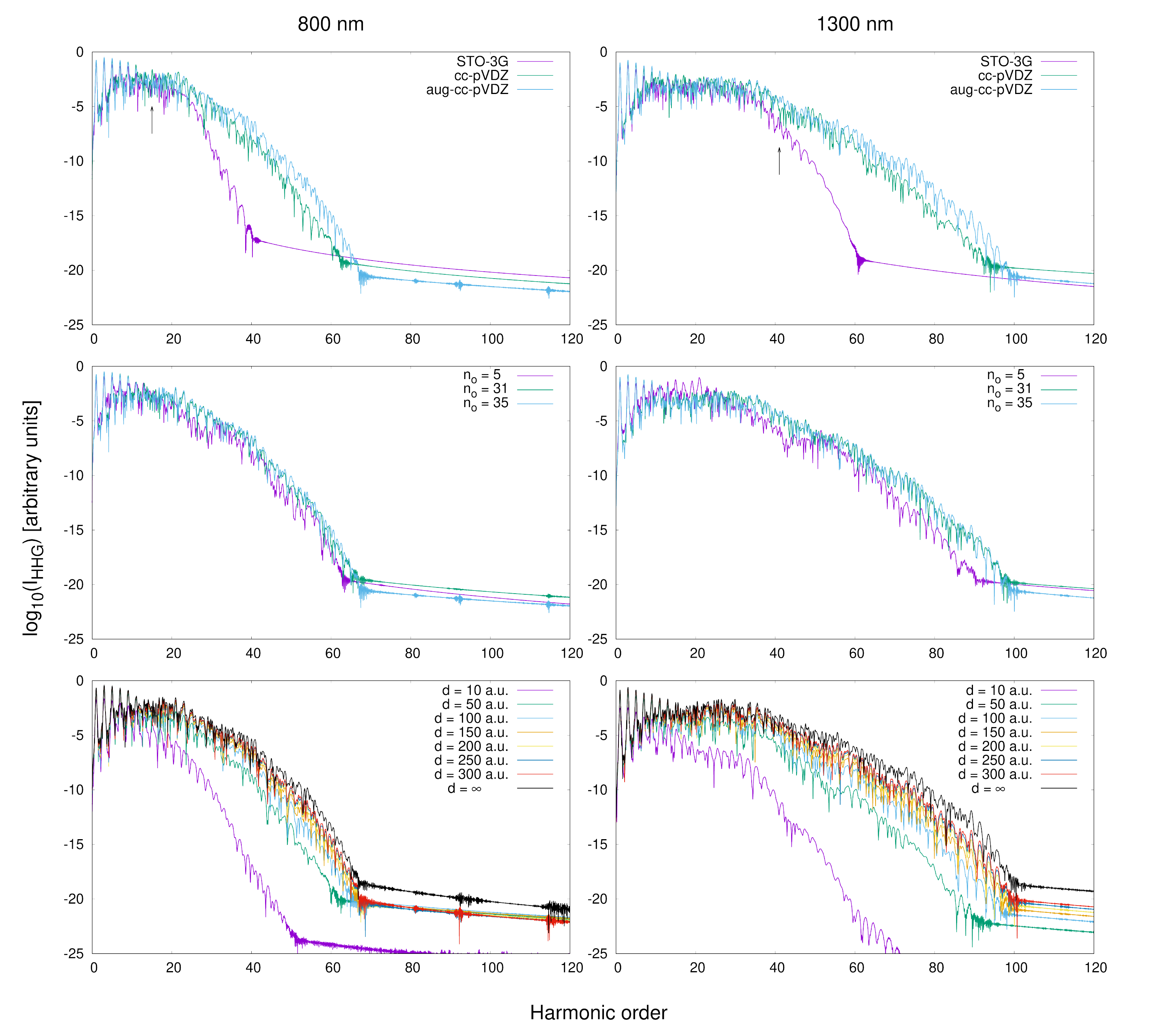}
  \caption{Results of the benchmark calculations of HHG in C\textsubscript{60} at the RT-TDTDA/B3LYP level of theory.
  Top row: comparison of the HHG spectra obtained using three different basis sets, with $n_o = 35$ and $d = 200$ a.u.
  Black vertical arrows denote the positions of the HHG cutoff as predicted by the 3SM.
  Middle row: comparison of the HHG spectra obtained in the aug-cc-pVDZ basis set using different numbers of active occupied MOs, with $d = 200$.
  For better clarity, only the results for $n_o = 5$, 31, and 35 are shown.
  Bottom row: comparison comparison of the HHG spectra obtained in the aug-cc-pVDZ basis set using different values of the $d$ parameter, with $n_o = 35$. }
  \label{fgr:convergence}
\end{figure}

The results of these preliminary calculations are presented in Fig. \ref{fgr:convergence}.
The two top plots show the comparison of three considered basis sets.
It can be seen that the STO-3G basis set provides a significantly different depiction of HHG in comparison to the Dunning basis sets.
Specifically, it predicts an abrupt drop in the HHG intensity near the cutoff energy predicted from the three step model, $E_{cut} = I_p + 3.17(E_0^2/4\omega_0^2)$, at both wavelengths.
In contrast, both cc-pVDZ and aug-cc-pVDZ spectra show a sizeable cutoff extension, accompanied by a more gradual reduction in the HHG intensity relative to the harmonic order.
This result is more in line with previous theoretical calculations of HHG in fullerenes and with the experimental observations.
Interestingly, the inclusion of the diffuse basis functions has a noticeably less pronounced effect on the description of HHG compared to the addition of the polarization functions.
This is evident as the spectra obtained in the cc-pVDZ and aug-cc-pVDZ basis sets do not differ significantly in terms of their overall shape.
The only distinctions are that the latter basis yields slightly higher peak intensities in the high-energy part of the spectrum and is capable of describing several additional peaks.
Since these differences are, nonetheless, noticeable, in the further calculations we use the largest aug-cc-pVDZ basis set.

The two middle plots in Fig. \ref{fgr:convergence} depict the influence of the number of the active occupied MOs $n_{o}$ on the obtained HHG spectra.
When increasing the active occupied space, we were able to achieve convergence at approximately $n_o \approx 30$, and, as can be seen, at both wavelengths the spectra obtained with $n_o = 31$ and $n_o = 35$ are nearly identical to each other.
Such an outcome is consistent with the somewhat intuitive reasoning that the $\pi$ band, consisting of 60 $p$-type orbitals of $sp^2$-hybridized carbon atoms, should be most strongly involved in the HHG process because the electrons occupying it can move freely throughout the entire molecule.
However, to ensure that the computed spectra are truly converged with respect to $n_o$, in the subsequent calculations we use the highest considered number of 35 occupied MOs.
In the plots, we also show the results for $n_o = 5$, which in our case is the closest to the SAE model.
Although not dramatically different from the converged ones, the spectra obtained with $n_o = 5$ are of noticeably inferior quality, with the decreased HHG intensity and artificial local minima in the high-energy region.
While some of the earliest HHG calculations on fullerenes based on the strong field approximation considered only electronic transitions from the HOMO level \cite{ciappina2007,ciappina2008}, our results indicate that this may not be the optimal strategy.

Finally, in the two bottom plots we present the dependence between the value of the $d$ parameter in the finite lifetime model and the computed HHG signal.
It is evident that applying the two lowest $d$ values leads to a significant reduction in HHG intensity and a decrease in the number of peaks, indicating that the absorption model interferes with the HHG process.
The calculated HHG response stabilizes at much higher values of $d \approx$ 150 bohr. The spectra obtained with $d = 150$, 200, 250, and 300 bohr exhibit consistent peak shapes and only marginal differences in overall intensity.
Therefore, we pick an intermediate value of $d = 200$ bohr for further calculations.
It is worth mentioning that the spectra shapes, once converged with respect to $d$, are not identical to the spectra shapes obtained without the finite lifetime model.
This implies that the absorber effectively elliminates the wavefunction refelctions arising from the basis set incompleteness, as intended, without hindering the HHG efficiency.

At this point we would like to note that unlike in calculations for smaller systems, we do not supplement the employed basis sets with additional diffuse functions designed to describe the electron dynamics in the continuum.
This is mainly dictated by practical considerations.
The rank of the $\mathbf{A}$ and $\mathbf{B}$ matrices constructed in the aug-cc-pVDZ basis set, using the full virtual space and active occupied space with $n_o = 35$, is equal to 42000.
Fully diagonalizing even larger matrices, though technically feasible, would demand significant time and resources.
To the best of our knowledge, this is already the highest number of excited states reported for use in RT-TDSEM-based calculations.
Moreover, in case of the C\textsubscript{60} fullerene, the aug-cc-pVDZ basis set is already on the verge of linear dependencies.
However, judging by the small differences between the cc-pVDZ and aug-cc-pVDZ spectra, we can infer that HHG in C\textsubscript{60} mainly occurs on the surface of the fullerene.
Therefore, addition of more diffuse functions is not necessary.
This is further evidenced by the high optimal $d$ value that greatly exceeds the maximum electron excursion distance under the considered laser conditions (equal to 11.6 bohr at 800 nm and 30.7 bohr at 1300 nm), even when extended by the largest internuclear distance in C\textsubscript{60} ($\approx$ 13.4 bohr).
The orbital ionization rate in the finite lifetime model, $\sqrt{2\epsilon_a}/d$, is interpreted as an inverse of the time required for an electron with kinetic energy $\epsilon_a$ to travel a distance $d$.
In this context, the parameter $d$ can be considered equivalent to the distance between the molecule and the starting point of the complex absorbing potential.
However, from Fig. \ref{fgr:convergence}, it is evident that setting $d$ similar to the electron excursion distance results in an overestimation of absorption.
This overestimation occurs due to the assignment of excessively short lifetimes to virtual MOs, akin to placing the complex absorbing potential too close to the molecule.
This effect implies that in C\textsubscript{60}, the proportionality between the virtual MO energy and the distance that an electron occupying this MO can travel within a given amount of time is no longer valid.
The probable cause is that even virtual MOs with relatively high energies are still localized in the vicinity of the molecular surface.
This indirectly confirms that fullerene HHG results from oscillations of the electron density within the molecule and cannot be fully described by the three-step model.

\subsection{Comparison of different exchange-correlation potentials}

In this section we compare the HHG spectra computed using the B3LYP functional with those obtained using the CAM-B3LYP functional and calculated at the HF level.
At this point, we adhere to the CIS/TDA approximation.
Let us briefly remind that while B3LYP and CAM-B3LYP share the same correlation functional, they differ in terms of the exchange functional.
B3LYP contains a fixed portion of 20\% HF exchange, which is known to lead to incorrect behavior of this functional at the long range.
Therefore, CAM-B3LYP has been proposed as a range-separated version of B3LYP, in which the percentage of HF exchange depends on interelectronic distance and varies from 19\% at the short range to 65\% at the long range \cite{yanai2004}.
While still not making it asymptotically correct, this considerably improves the long-range behavior of the functional.

\begin{figure}
  \includegraphics[width=0.99\linewidth]{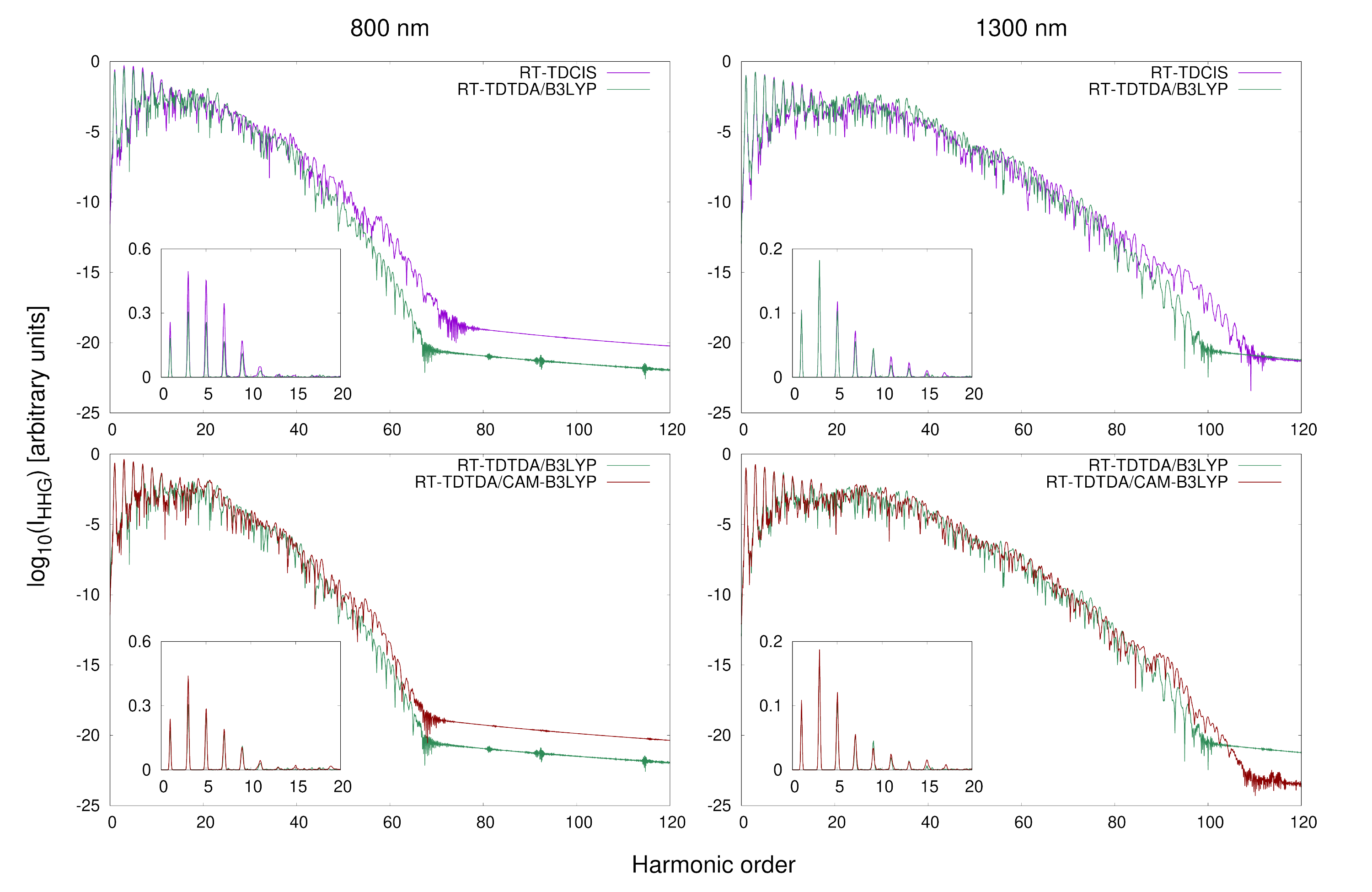}
  \caption{Comparison of the HHG spectra of C\textsubscript{60} calculated in the aug-cc-pVDZ basis set at the RT-TDCIS, RT-TDTDA/B3LYP, and RT-TDTDA/CAM-B3LYP level of theory. The insets show the lowest-energy regions of the spectra, plotted in linear scale}
  \label{fgr:vxc}
\end{figure}

The HHG spectra obtained from RT-TDCIS, RT-TDTDA/B3LYP, and RT-TDTDA/CAM-B3LYP are shown on Fig. \ref{fgr:vxc}.
All three methods provide spectra of comparable quality in terms of overall shape and intensity.
However, we can distinguish two regions where certain systematic differences can be observed depending on $f_{xc}$ used.
The first of them is the lowest-energy part of the spectrum, where individual peaks are most clearly visible.
It can be seen that RT-TDCIS predicts significantly higher intensities of peaks up to 20th harmonic order compared to both variants of RT-TDTDA, especially at 800 nm.
In our opinion, this part of the spectrum is primarily infuenced by the short-range interactions, specifically the short-range correlation.
The peaks with the lowest energy levels represent transitions to and from the least energetic excited states, during which the excited electron remains in close proximity to the molecule.
Therefore, the dynamical correlation effects are expected be most prominent in this region.
Since the correlation part of B3LYP is the same as of CAM-B3LYP, the peak intensities in this part of the spectrum predicted by these two functionals are much more similar to each other.
We have already discussed the effect of decreased HHG intensity due to correlation effects in our works on smaller systems \cite{wozniak2022,wozniak2023}.

\begin{figure}
  \includegraphics[width=0.50\linewidth]{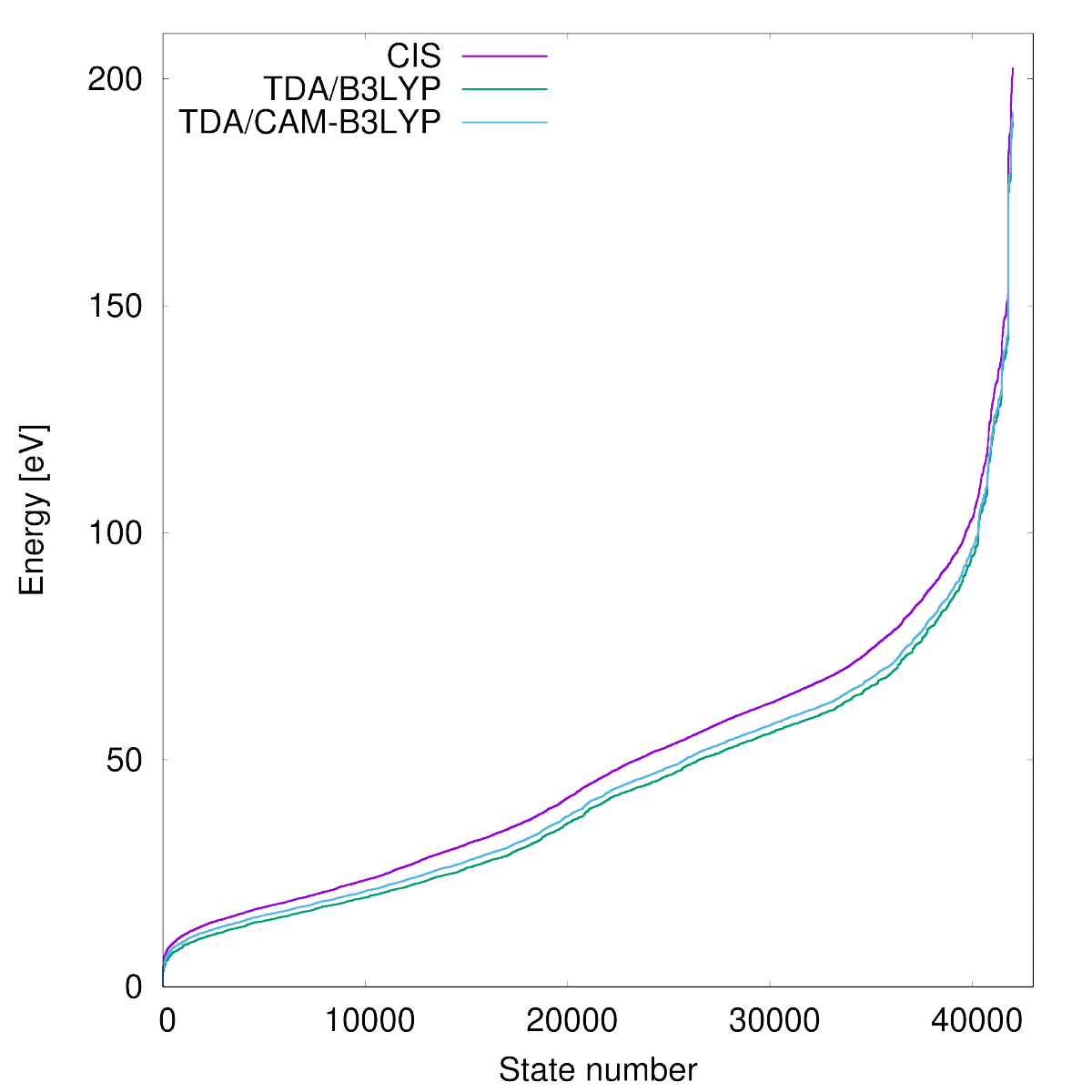}
  \caption{Distributions of the excitation energies obtained using CIS, TDA/B3LYP and TDA/CAM-B3LYP in the aug-cc-pVDZ basis set}
  \label{fgr:states_xc}
\end{figure}

The second region encompasses the highest-energy part of the spectrum, extending beyond the 40th harmonic at 800 nm and beyond the 80th harmonic at 1300 nm.
By analogy, we suspect that the description of this region is mainly governed by the long-range interactions.
It can be noticed that the intensity of the spectrum ``tail" is more or less proportional to the amount of the HF exchange in the exchange-correlation functional.
RT-TDCIS, which provides an asymptotically correct description of exchange effects, predicts the most intense peaks in this region.
In contrast, RT-TDTDA/B3LYP, which is the least accurate at the long-range limit, predicts peaks of the lowest intensity.
Additionally, at 1300 nm, both RT-TDCIS and RT-TDTDA/CAM-B3LYP are able to describe a few more peaks than RT-TDTDA/B3LYP.
A similar pattern can also be seen when comparing the distributions of excited states  obtained using different exchange-correlation potentials (Fig.~\ref{fgr:states_xc}).
Introducing correlation by replacing the HF potential with B3LYP lowers nearly all excitation energies, but the correction of the long-range exchange has a counteracting effect, causing a slight shift back towards higher values.

\subsection{Influence of deexcitation effects on the HHG response}

In Fig. 5, we compare the HHG spectra obtained in the TDA/CIS approximation with those obtained using the full RPA framework, separately for each of the three considered correlation-exchange potentials.
Including the $\mathbf{B}$ matrix when solving the linear-response problem has a much smaller impact on the description of HHG than the choice of the exchange-correlation potential.
In all cases the spectra computed using RPA and TDA/CIS eigenstates nearly overlap with each other.
The only systematic effect is that RT-TDRPA predicts a slightly lower HHG intensity across the entire spectrum, which is more visible at 800 nm.
Also, this effect is subtly more pronounced for RT-TDRPA/DFT than for RT-TDRPA/HF.
The only exception can be observed for CAM-B3LYP, where the backgroud level of the RT-TDRPA spectrum is upshifted compared to the RT-TDTDA spectrum, resulting in a reduction of the number of described peaks in the highest-energy region.

\begin{figure}
  \includegraphics[width=0.99\linewidth]{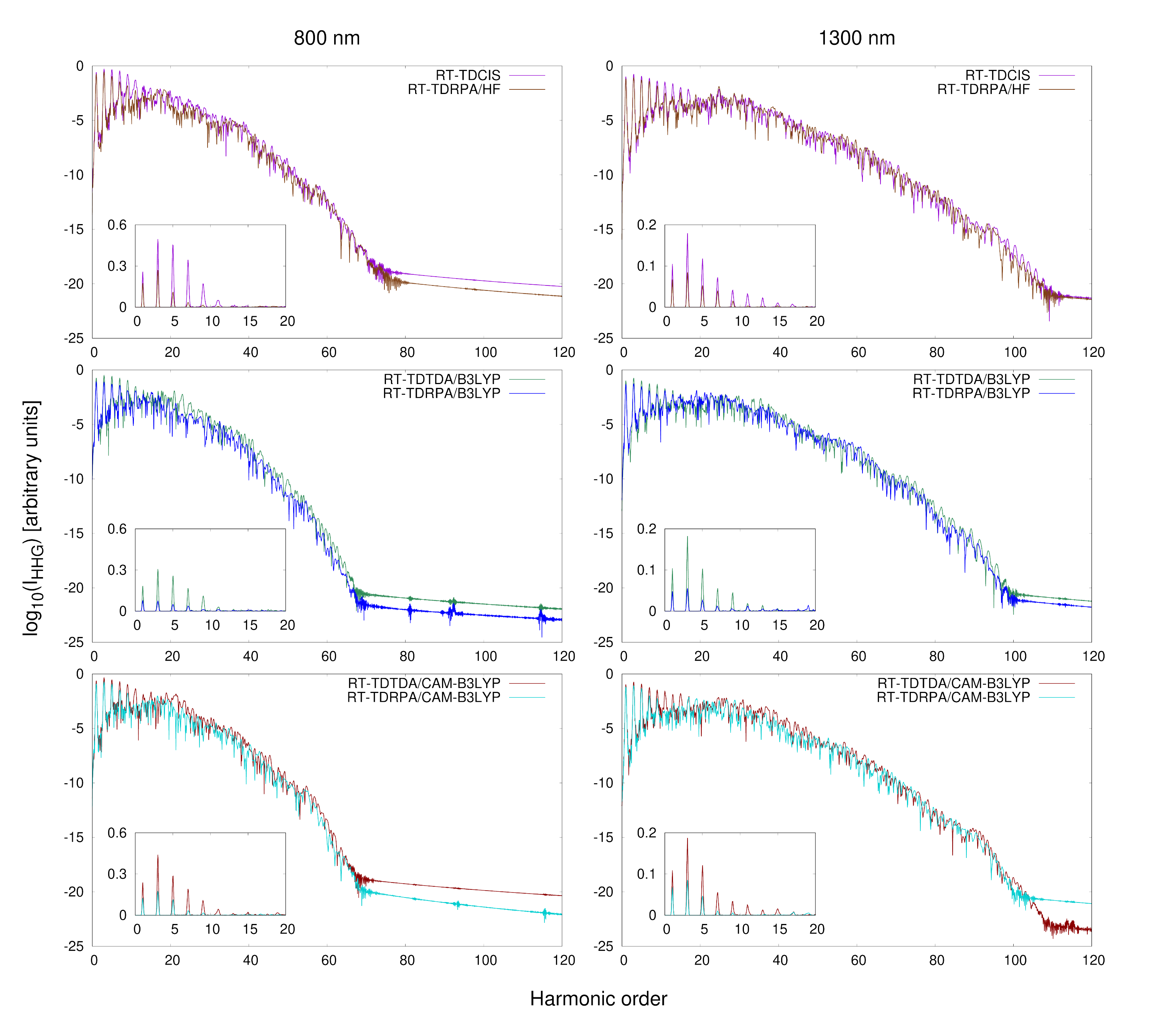}
  \caption{Comparison of the HHG spectra of C\textsubscript{60} calculated in the aug-cc-pVDZ basis set: RT-TDCIS vs RT-TDRPA/HF (top row), RT-TDTDA/B3LYP vs RT-TDRPA/B3LYP (middle row), and RT-TDTDA/CAM-B3LYP vs RT-TDRPA/CAM-B3LYP (bottom row). The insets show the lowest-energy regions of the spectra, plotted in linear scale}
  \label{fgr:rpa}
\end{figure}

\begin{figure}
  \includegraphics[width=0.50\linewidth]{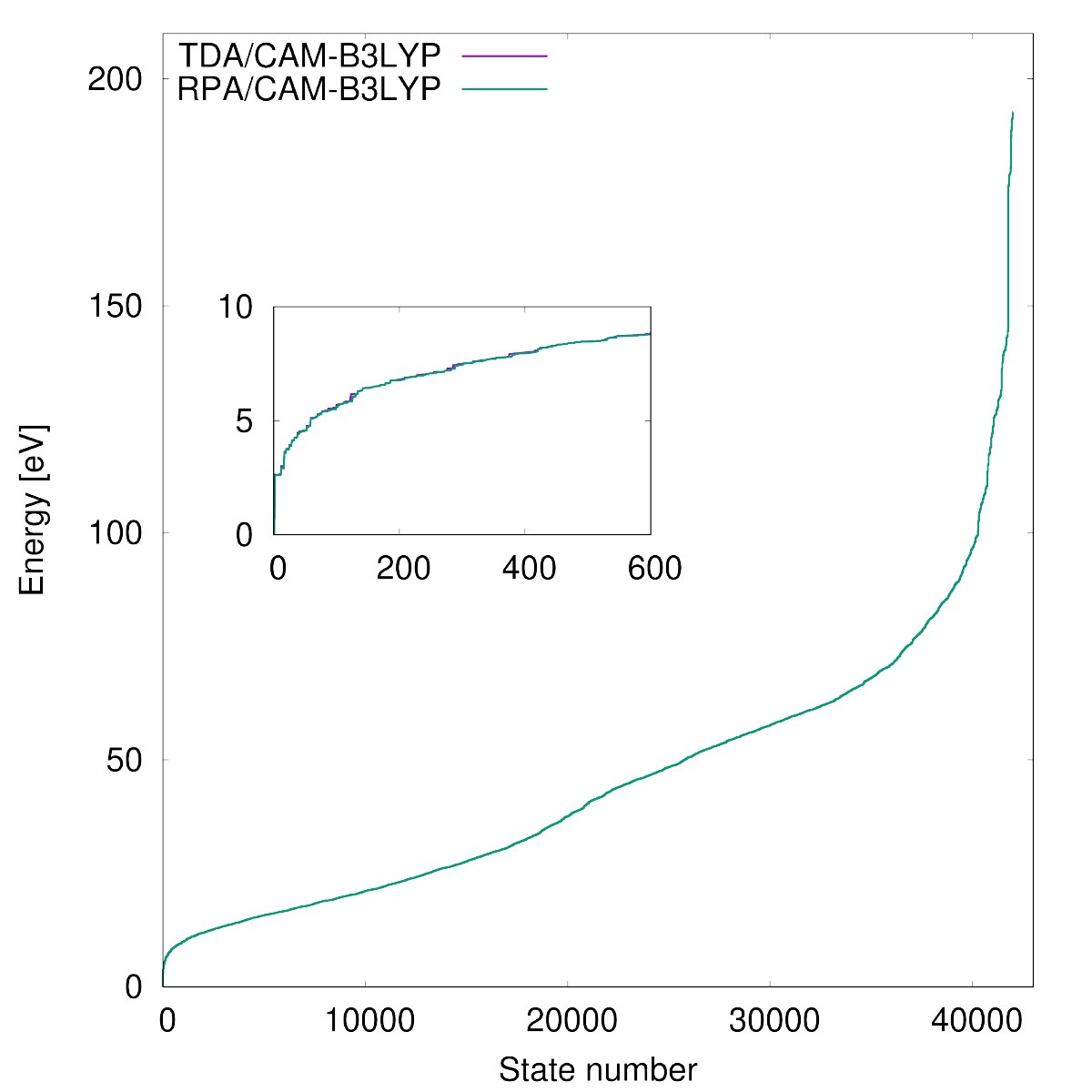}
  \caption{Distributions of the excitation energies obtained using TDA/CAM-B3LYP and RPA/CAM-B3LYP in the aug-cc-pVDZ basis set. The inset shows the zoom of 600 lowest excitation energies}
  \label{fgr:states_rpa}
\end{figure}

Interestingly, RT-TDRPA most significantly reduces the intensity of HHG in the lowest-energy region of the spectra, where the intensities of first few peaks are 2-4 times lower compared to RT-TDTDA and RT-TDCIS.
Although we explained the similar differences between RT-TDTDA and RT-TDCIS spectra by the correlation effects, such an explanation may not be appropriate in this case.
While RPA is recognized to be a correlated method for the ground state, it is a matter of debate whether this property applies to excited states as well.
Recently, Berkelbach has shown that LR-TDHF is equivalent to a variant of EOM-CCD, in which the CCD ground state is taken as a reference state, but only single excitations are considered when solving the equations of motion \cite{berkelbach2018}.
This implies that the RPA excitation energies do not include any additional correlation effects beyond those already accounted for in the exchange-correlation kernel, especially given that in all flavors of RT-TDRPA we use a single-reference ground state.
We reach a similar conclusion when analysing the excited state distributions obtained from RPA and TDA/CIS calculations, which are almost identical to each other (Fig. \ref{fgr:states_rpa}).
We believe that the differences between the peak intensities may instead come from the differences in the dipole moments and dipole transition moments between TDA/CIS and RPA states.
It is known that LR-TDHF and LR-TDDFT obey the Thomas-Reiche-Kuhn sum rule, unlike CIS and TDA \cite{dreuw2005}.
Therefore, one should anticipate an improved description of the ground-to-excited transitions in the former two methods.
On the other hand, we calculate the excited-to-excited transition moments between LR-TDHF and LR-TDDFT using the approximate formula \eqref{eerpa}, whereas the analogous Eq. \eqref{eecis} is a valid expression for CIS and TDA eigenstates.
Due to these two potential sources of errors, it is not possible to definitively determine which approach, with or without including the $\mathbf{B}$ matrix, allows for a better description of electron dynamics.

\begin{figure}
  \includegraphics[width=0.50\linewidth]{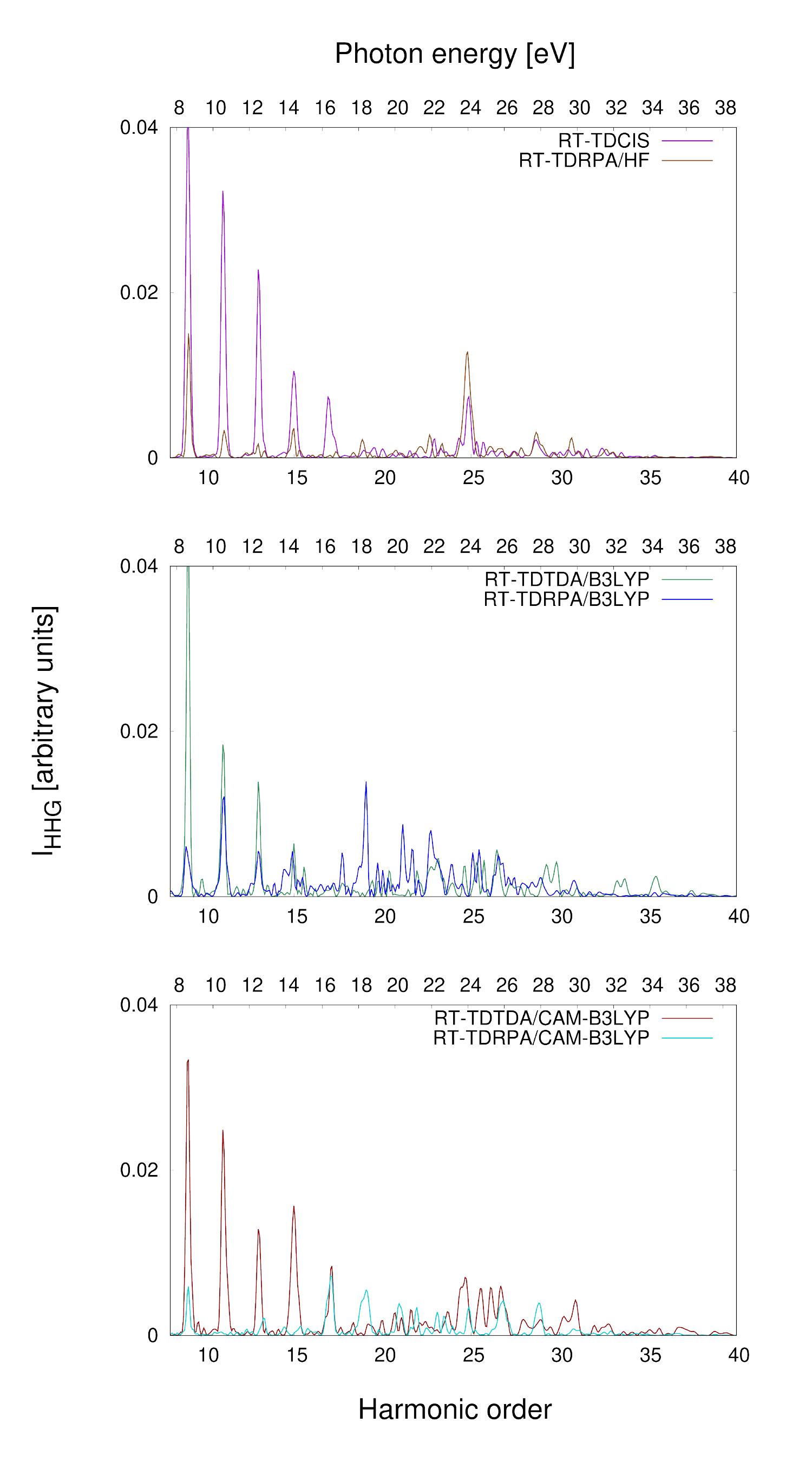}
  \caption{Excerpts of the spectra computed at 1300 nm from Fig. \ref{fgr:rpa}, covering the region in which the giant dipole resonance is expected to occur}
  \label{fgr:gdr}
\end{figure}

When analysing the computed HHG signals, we noticed that some of the spectra calculated at 1300 nm display a local maximum of intensity between 15th and 30th harmonic order.
At this wavelength, this region corresponds to photon energies of 14-29 eV.
This is the energy range in which the giant dipole resonance (GDR) can be observed in the photoionization spectra of the C\textsubscript{60} fullerene \cite{gensterblum1991,hertel1992,yoo1992,ju1993,scully2005}.
The occurence of such dipole resonances in known to enhance the harmonic intensity \cite{ganeev2011}.
Therefore, we take a closer look at this region of the spectra in Fig. \ref{fgr:gdr}.
Surprisingly, the GDR is most clearly visible in the RT-TDCIS and RT-TDRPA/HF spectra, where a single peak at 25th harmonic order (corresponding to approximately 23.8 eV) is significantly enhanced.
RT-TDTDA/CAM-B3LYP provides a not too dissimilar picture, as an enhancement of a group of peaks between 24th and 27th harmonic order (22.9-25.7 eV) can be observed.
Both of these results are in reasonable agreement with experimental works that predict the GDR maximum at 20-22 eV for C\textsubscript{60} and 21-24 eV for C\textsubscript{60}\textsuperscript{+} \cite{hertel1992,yoo1992,scully2005}.
RT-TDRPA/B3LYP predicts the strongest enhancement of 19th peak, corresponding to a somewhat lower energy of 18.1 eV.
Finally, practically no enhancement can be seen in the RT-TDTDA/B3LYP and RT-TDRPA/CAM-B3LYP spectra.
This simple test allows us to draw a conclusion that the full RPA treatment of the excited states may indeed result in a less accurate description of the electron dynamics in C\textsubscript{60}, at least when a Kohn-Sham determinant is employed as a reference, as evidenced by the case of CAM-B3LYP calculations.
Furthermore, the fact that B3LYP is unable to predict the GDR at the RT-TDTDA level and provides its incorrect position at the RT-TDRPA level may indicate that long-range exchange effects are particularly important for describing the global resonances in C\textsubscript{60}.
Unfortunately, we are unable to detect the GDR-related enhancement at 800 nm, as at this wavelength it should be located at lower harmonic orders, where the overall HHG intensity is several orders of magnitude higher.

\subsection{Semiempirical HHG calculations}

\begin{figure}
  \includegraphics[width=0.99\linewidth]{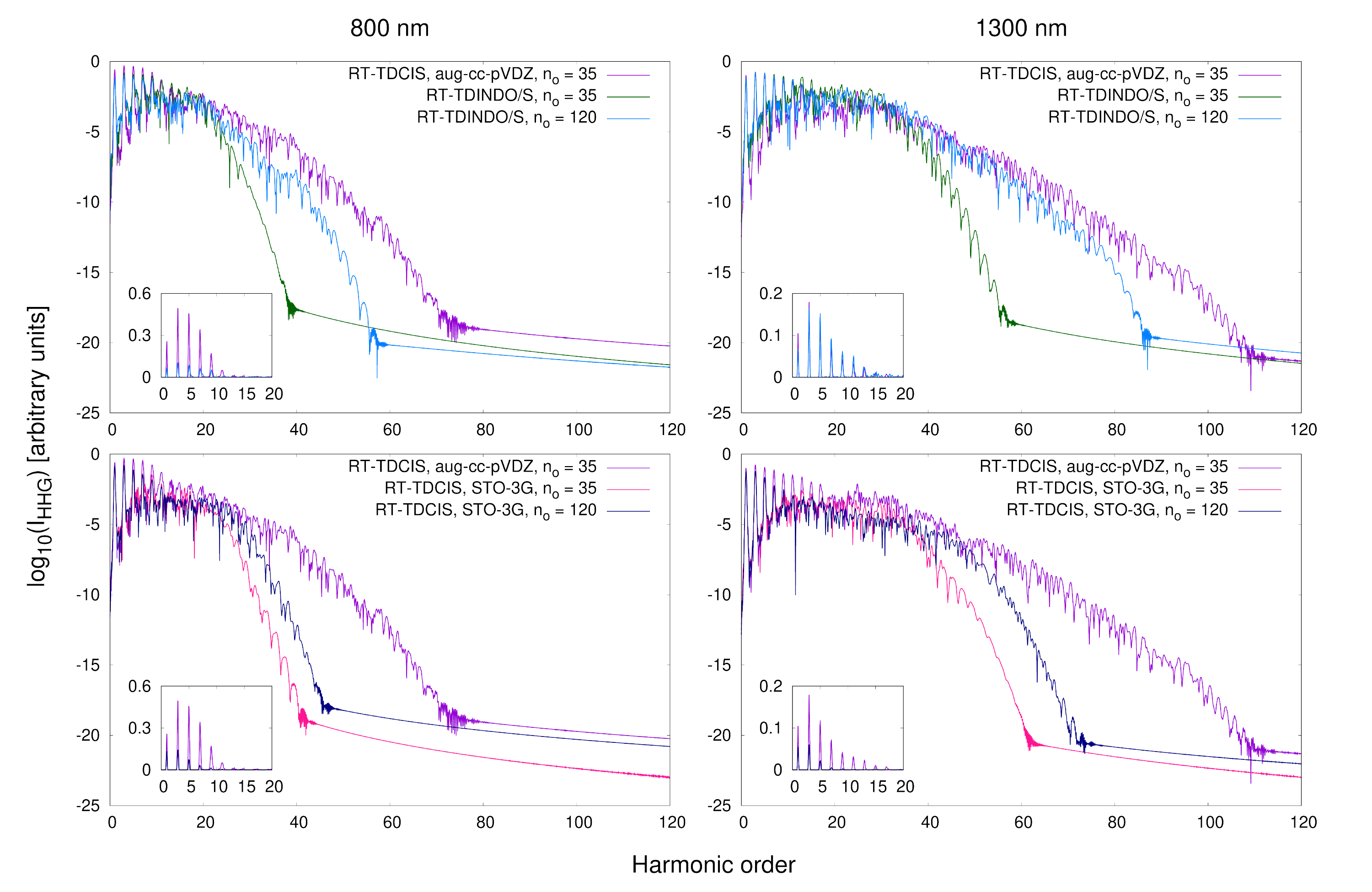}
  \caption{Top row: comparison of the HHG spectra computed using RT-TDCIS in the aug-cc-pVDZ basis set, and using RT-TDINDO/S within two different active occupied spaces. Bottom row: comparison of the HHG spectra computed using RT-TDCIS in the aug-cc-pVDZ basis set, and in the STO-3G minimal basis set within two different active occupied spaces}
  \label{fgr:indo}
\end{figure}

Finally, in this section we analyze the results obtained from the RT-TDINDO/S propagations.
As mentioned earlier, in RT-TDINDO/S simulations we employ the Slater minimal basis set compatible with semiempirical methods, while the size of the active occupied space and the value of the $d$ parameter remains the same as in the \textit{ab initio} and DFT-based calculations.
Since INDO/S can be considered a semiempirical counterpart of CIS, a natural choice is to compare the computed spectra with RT-TDCIS ones.
Such a comparison is shown in two upper plots in Fig.~\ref{fgr:indo}, where the purple curves represent RT-TDCIS results and the green curves represent RT-TDINDO/S results.
RT-TDINDO/S successfully reproduces most of the peaks in the low-energy portion of the spectra.
This is a commendable result, considering the approximate nature of the INDO/S Hamiltonian.
However, the picture it provides at higher energies closely resembles that of RT-TDTDA/B3LYP in the minimal basis set (shown in Fig. \ref{fgr:convergence}), with a sudden decrease in the HHG intensity at too low energies.
Nevertheless, before we deem RT-TDINDO/S unsuitable for HHG modeling, we must recall that the INDO/S Hamiltonian was parameterized to replicate CIS excitation energies in an already truncated active orbital space.
Therefore, further constraining the active occupied space of INDO/S itself may yield suboptimal results.
Having that in mind, we perform a second series of calculations, this time allowing excitations from all 120 INDO/S occupied MOs (light-blue curves on Fig.~\ref{fgr:indo}).
This hugely improves the HHG depiction in the high-energy range.
The positions of the last described peaks are upshifted from 37th to 57th harmonic order at 800 nm and from 57th to 87th harmonic order at 1300 nm.
Additionally, the overall intensity of HHG now aligns more closely with that predicted by RT-TDCIS over a much broader energy range.
Interestigly, RT-TDINO/S performs notably better at 1300 nm, where a good agreement with the RT-TDCIS curve is observed up to approximately 50th harmonic order and the intensities of the lowest-energy peaks are very closely reproduced.
At 800 nm, RT-TDINDO/S predicts two regions where the intensity of high-harmonic generation (HHG) significantly decreases, the first one around 23th harmonic order and the second around 43rd harmonic order, leading to larger discrepancies with the RT-TDCIS predictions.

The above results give prompt a question about whether the observed improvement of RT-TDINDO/S performance is genuinely attributable to the effectiveness of the semiempirical Hamiltonian, or is merely a consequence of expanding the configurational space.
To address it, we conduct analogous calculations at the RT-TDCIS using the STO-3G basis set, which is our closest analogue to the INDO/S Slater minimal basis set.
Here, we also use two different active occupied spaces, with $n_o = 35$ and $n_o = 120$.
The results are presented on the two bottom plots of Fig.~\ref{fgr:indo}.
While an increase in $n_o$ modestly extends the cutoff at both intensities, the magnitude of this extension is nowhere near what is observed in the RT-TDINDO/S spectra.
The position of the last described peak is shifted upwards by about 5 harmonic orders at 800 nm and by about 12 harmonic orders at 1300 nm.

\begin{figure}
  \includegraphics[width=0.99\linewidth]{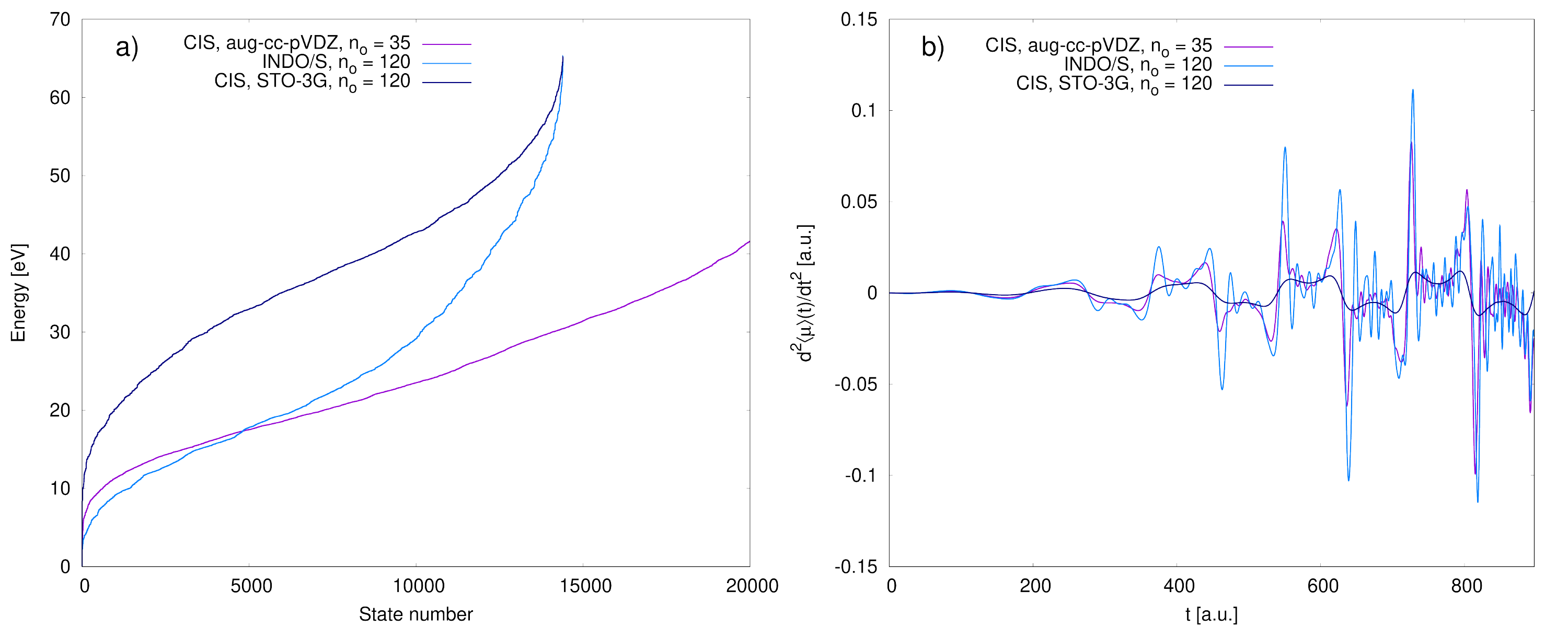}
  \caption{a) Distributions of excitation energies calculated with CIS in the aug-cc-pVDZ and STO-3G basis sets and with INDO/S b) Time-resolved dipole acceleration during the first five optical cycles of the applied laser pulse of 1300 nm, obtained from the RT-TDCIS propagations in the aug-cc-pVDZ and STO-3G basis sets and from the RT-TDINDO/S propagation. Calculation involving CIS in the STO-3G basis set and INDO/S employed an increased active occupied space with $n_o = 120$}
  \label{fgr:states_indo}
\end{figure}

The superior performance of RT-TDINDO/S is also evident when comparing the excitation energies obtained from the respective linear-response equations (Fig.~\ref{fgr:states_indo}a).
The distribution of excited states obtained using INDO/S, particularly those below 20 eV, relatively closely matches the one calculated with CIS in the Dunning basis set.
In contrast, CIS in the minimal basis set significantly overestimates the excitation energies.
It is worth highlighting that such a notable discrepancy can be observed despite the fact that exactly the same number of excited states is available in both Slater and Gaussian minimal basis set, totaling 14400.
These disparities directly impact the time-resolved optical response, as seen on Fig.~\ref{fgr:states_indo}b.
RT-TDINDO/S reasonably reproduces the oscillations in the dipole acceleration, with a moderate overestimation of amplitudes but accurate preservation of their temporal positions.
This overestimation can be attributed to the underestimation of the excitation energies below $\sim$ 15 eV, resulting in an overpopulation of the respective excited states.
On the other hand, the curve corresponding to CIS in the STO-3G basis set exhibits a smoother, less oscillatory character due to the limited availability of energetically accessible electronic transitions states.
This translates to a reduction in the number of harmonic peaks in the HHG spectrum.

\section{Conclusion}

In this study we conducted and analyzed a series of calculations of the HHG spectra of the C\textsubscript{60} fullerene in the non-perturbational regime, employing various quantum chemical methods.
These include the real-time time-dependent counterparts of CIS, TDA and RPA based on either the Hartree-Fock or Kohn-Sham determinant, as well as the semiempirical INDO/S method.

The main conclusion that can be drawn from this work is that \textit{ab initio} and DFT-based methods coupled to Gaussian bases can be successfully applied to model strong-field processes in nanoscale systems.
Naturally, these calculations are much more resource-intensive compared to those on atoms and smaller molecules, as most of the performed propagations involved expanding the time-dependent wavefunction in several tens of thousands of states.
Nevertheless, owing to the linearity of the real-time propagation equations, the computational costs remain reasonable, and represent a modest price for an effectively all-electron picture of the laser-driven dynamics.
Our calculations on C\textsubscript{60} correctly prectict that the high harmonic generation in fullerenes primarily arises from the oscillations of the electronic denisty at the molecular surface, with a predominant contribution from the 60 $\pi$ electrons to this process.
We also demonstrated that RT-TDSEM are able to detect more subtle features of the attosecond processes, such as the giant dipole resonance.

Our results indicate that the quantum chemical description of the electron dynamics primarily depends on the chosen basis set and the active orbital space employed.
The influence of the exchange-correlation potential, on the other hand, is relatively minor.
Nevertheless, we were able to identify some features of the spectra that can be attributed to both correlation and exchange effects.
Based on our findings, we can recommend the use of range-separated DFT functionals in the calculations involving systems with a significant role of dynamical correlation.
In our case, CAM-B3LYP successfully combines the good approximation to the short-range correlation of B3LYP with the proper treatment of the long-range exchange characteristic to HF-based methods.
On the other hand, a full RPA description of the excited electronic states provides practically no improvement over CIS, and may, in fact, lead to inferior results compared to TDA based on the Kohn-Sham reference.

Finally, we have demonstrated that the semiempricial INDO/S Hamiltonian, traditionally recognized for offering reasonably accurate approximations to the excitation energies, can also be successfully used for modeling strong-field processes in large systems.
The performance of RT-TDINDO/S surpasses that of RT-TDCIS and RT-TDA in a basis set of comparable size, yielding results in qualitative agreement with all-electron approaches employing larger basis sets, particularly at lower harmonic orders and for relatively long laser wavelengths.
This positions it as a potentially valuable simulation framework for studying even larger systems, which are gaining increasing interest in attosecond science but exceed the capabilities of the ab \textit{ab initio} and DFT-based methods.
The applicability of RT-TDINDO/S will be further explored in future works.

\begin{acknowledgement}

This work was supported by the Polish National Science Centre (NCN) through Grant No. 2017/25/B/ST4/02698. The calculations have been carried out using resources provided by Wroclaw Centre for Networking and Supercomputing, Grant No. 567, and by University of Arizona Research Computing, courtesy of Prof. Ludwik Adamowicz.

\end{acknowledgement}




\bibliography{main}

\providecommand{\latin}[1]{#1}
\makeatletter
\providecommand{\doi}
  {\begingroup\let\do\@makeother\dospecials
  \catcode`\{=1 \catcode`\}=2 \doi@aux}
\providecommand{\doi@aux}[1]{\endgroup\texttt{#1}}
\makeatother
\providecommand*\mcitethebibliography{\thebibliography}
\csname @ifundefined\endcsname{endmcitethebibliography}  {\let\endmcitethebibliography\endthebibliography}{}
\begin{mcitethebibliography}{144}
\providecommand*\natexlab[1]{#1}
\providecommand*\mciteSetBstSublistMode[1]{}
\providecommand*\mciteSetBstMaxWidthForm[2]{}
\providecommand*\mciteBstWouldAddEndPuncttrue
  {\def\EndOfBibitem{\unskip.}}
\providecommand*\mciteBstWouldAddEndPunctfalse
  {\let\EndOfBibitem\relax}
\providecommand*\mciteSetBstMidEndSepPunct[3]{}
\providecommand*\mciteSetBstSublistLabelBeginEnd[3]{}
\providecommand*\EndOfBibitem{}
\mciteSetBstSublistMode{f}
\mciteSetBstMaxWidthForm{subitem}{(\alph{mcitesubitemcount})}
\mciteSetBstSublistLabelBeginEnd
  {\mcitemaxwidthsubitemform\space}
  {\relax}
  {\relax}

\bibitem[Agostini and DiMauro(2004)Agostini, and DiMauro]{agostini2004}
Agostini,~P.; DiMauro,~L.~F. The physics of attosecond light pulses. \emph{Rep. Prog. Phys.} \textbf{2004}, \emph{67}, 813--855\relax
\mciteBstWouldAddEndPuncttrue
\mciteSetBstMidEndSepPunct{\mcitedefaultmidpunct}
{\mcitedefaultendpunct}{\mcitedefaultseppunct}\relax
\EndOfBibitem
\bibitem[Krausz and Ivanov(2009)Krausz, and Ivanov]{krausz2009}
Krausz,~F.; Ivanov,~M. Attosecond physics. \emph{Rev. Mod. Phys.} \textbf{2009}, \emph{81}, 163\relax
\mciteBstWouldAddEndPuncttrue
\mciteSetBstMidEndSepPunct{\mcitedefaultmidpunct}
{\mcitedefaultendpunct}{\mcitedefaultseppunct}\relax
\EndOfBibitem
\bibitem[Vozzi \latin{et~al.}(2011)Vozzi, Negro, Calegari, Sansone, Nisoli, De~Silvestri, and Stagira]{vozzi2011}
Vozzi,~C.; Negro,~M.; Calegari,~F.; Sansone,~G.; Nisoli,~M.; De~Silvestri,~S.; Stagira,~S. Generalized molecular orbital tomography. \emph{Nat. Phys.} \textbf{2011}, \emph{7}, 822--826\relax
\mciteBstWouldAddEndPuncttrue
\mciteSetBstMidEndSepPunct{\mcitedefaultmidpunct}
{\mcitedefaultendpunct}{\mcitedefaultseppunct}\relax
\EndOfBibitem
\bibitem[Salières \latin{et~al.}(2012)Salières, Maquet, Haessler, Caillat, and Taïeb]{salieres2012}
Salières,~P.; Maquet,~A.; Haessler,~S.; Caillat,~J.; Taïeb,~R. Imaging orbitals with attosecond and Ångström resolutions: toward attochemistry? \emph{Rep. Prog. Phys.} \textbf{2012}, \emph{75}, 062401\relax
\mciteBstWouldAddEndPuncttrue
\mciteSetBstMidEndSepPunct{\mcitedefaultmidpunct}
{\mcitedefaultendpunct}{\mcitedefaultseppunct}\relax
\EndOfBibitem
\bibitem[Peng \latin{et~al.}(2019)Peng, Marceau, and Villeneuve]{peng2019}
Peng,~P.; Marceau,~C.; Villeneuve,~D.~M. Attosecond imaging of molecules using high harmonic spectroscopy. \emph{Nat. Rev. Phys.} \textbf{2019}, \emph{1}, 144--155\relax
\mciteBstWouldAddEndPuncttrue
\mciteSetBstMidEndSepPunct{\mcitedefaultmidpunct}
{\mcitedefaultendpunct}{\mcitedefaultseppunct}\relax
\EndOfBibitem
\bibitem[W{\"o}rner \latin{et~al.}(2010)W{\"o}rner, Bertrand, Kartashov, Corkum, and Villeneuve]{worner2010a}
W{\"o}rner,~H.~J.; Bertrand,~J.~B.; Kartashov,~D.~V.; Corkum,~P.~B.; Villeneuve,~D.~M. Following a chemical reaction using high-harmonic interferometry. \emph{Nature} \textbf{2010}, \emph{466}, 604--607\relax
\mciteBstWouldAddEndPuncttrue
\mciteSetBstMidEndSepPunct{\mcitedefaultmidpunct}
{\mcitedefaultendpunct}{\mcitedefaultseppunct}\relax
\EndOfBibitem
\bibitem[Baykusheva \latin{et~al.}(2019)Baykusheva, Zindel, Svoboda, Bommeli, Ochsner, Tehlar, and W{\"o}rner]{baykusheva2019}
Baykusheva,~D.; Zindel,~D.; Svoboda,~V.; Bommeli,~E.; Ochsner,~M.; Tehlar,~A.; W{\"o}rner,~H.~J. Real-time probing of chirality during a chemical reaction. \emph{Proc. Natl. Acad. Sci. U.S.A} \textbf{2019}, \emph{116}, 23923--23929\relax
\mciteBstWouldAddEndPuncttrue
\mciteSetBstMidEndSepPunct{\mcitedefaultmidpunct}
{\mcitedefaultendpunct}{\mcitedefaultseppunct}\relax
\EndOfBibitem
\bibitem[Torres \latin{et~al.}(2010)Torres, Siegel, Brugnera, Procino, Underwood, Altucci, Velotta, Springate, Froud, Turcu, Patchkovskii, Ivanov, Smirnova, and Marangos]{torres2010}
Torres,~R.; Siegel,~T.; Brugnera,~L.; Procino,~I.; Underwood,~J.~G.; Altucci,~C.; Velotta,~R.; Springate,~E.; Froud,~C.; Turcu,~I. C.~E. \latin{et~al.}  Revealing molecular structure and dynamics through high-order harmonic generation driven by mid-IR fields. \emph{Phys. Rev. A} \textbf{2010}, \emph{81}, 051802(R)\relax
\mciteBstWouldAddEndPuncttrue
\mciteSetBstMidEndSepPunct{\mcitedefaultmidpunct}
{\mcitedefaultendpunct}{\mcitedefaultseppunct}\relax
\EndOfBibitem
\bibitem[Wong \latin{et~al.}(2011)Wong, Brichta, Spanner, Patchkovskii, and Bhardwaj]{wong2011}
Wong,~M. C.~H.; Brichta,~J.-P.; Spanner,~M.; Patchkovskii,~S.; Bhardwaj,~V.~R. High-harmonic spectroscopy of molecular isomers. \emph{Phys. Rev. A} \textbf{2011}, \emph{84}, 051403\relax
\mciteBstWouldAddEndPuncttrue
\mciteSetBstMidEndSepPunct{\mcitedefaultmidpunct}
{\mcitedefaultendpunct}{\mcitedefaultseppunct}\relax
\EndOfBibitem
\bibitem[W\"orner \latin{et~al.}(2009)W\"orner, Niikura, Bertrand, Corkum, and Villeneuve]{worner2009}
W\"orner,~H.~J.; Niikura,~H.; Bertrand,~J.~B.; Corkum,~P.~B.; Villeneuve,~D.~M. Observation of Electronic Structure Minima in High-Harmonic Generation. \emph{Phys. Rev. Lett.} \textbf{2009}, \emph{102}, 103901\relax
\mciteBstWouldAddEndPuncttrue
\mciteSetBstMidEndSepPunct{\mcitedefaultmidpunct}
{\mcitedefaultendpunct}{\mcitedefaultseppunct}\relax
\EndOfBibitem
\bibitem[Bruner \latin{et~al.}(2016)Bruner, Mašín, Negro, Morales, Brambila, Devetta, Faccialà, Harvey, Ivanov, Mairesse, Patchkovskii, Serbinenko, Soifer, Stagira, Vozzi, Dudovich, and Smirnova]{bruner2016}
Bruner,~B.~D.; Mašín,~Z.; Negro,~M.; Morales,~F.; Brambila,~D.; Devetta,~M.; Faccialà,~D.; Harvey,~A.~G.; Ivanov,~M.; Mairesse,~Y. \latin{et~al.}  Multidimensional high harmonic spectroscopy of polyatomic molecules: detecting sub-cycle laser-driven hole dynamics upon ionization in strong mid-IR laser fields. \emph{Faraday Discuss.} \textbf{2016}, \emph{194}, 369--405\relax
\mciteBstWouldAddEndPuncttrue
\mciteSetBstMidEndSepPunct{\mcitedefaultmidpunct}
{\mcitedefaultendpunct}{\mcitedefaultseppunct}\relax
\EndOfBibitem
\bibitem[W\"orner \latin{et~al.}(2010)W\"orner, Bertrand, Hockett, Corkum, and Villeneuve]{worner2010b}
W\"orner,~H.~J.; Bertrand,~J.~B.; Hockett,~P.; Corkum,~P.~B.; Villeneuve,~D.~M. Controlling the Interference of Multiple Molecular Orbitals in High-Harmonic Generation. \emph{Phys. Rev. Lett.} \textbf{2010}, \emph{104}, 233904\relax
\mciteBstWouldAddEndPuncttrue
\mciteSetBstMidEndSepPunct{\mcitedefaultmidpunct}
{\mcitedefaultendpunct}{\mcitedefaultseppunct}\relax
\EndOfBibitem
\bibitem[Shiner \latin{et~al.}(2011)Shiner, Schmidt, Trallero-Herrero, W{\"o}rner, Patchkovskii, Corkum, Kieffer, L{\'e}gar{\'e}, and Villeneuve]{shiner2011}
Shiner,~A.; Schmidt,~B.; Trallero-Herrero,~C.; W{\"o}rner,~H.~J.; Patchkovskii,~S.; Corkum,~P.~B.; Kieffer,~J.; L{\'e}gar{\'e},~F.; Villeneuve,~D. Probing collective multi-electron dynamics in xenon with high-harmonic spectroscopy. \emph{Nat. Phys.} \textbf{2011}, \emph{7}, 464--467\relax
\mciteBstWouldAddEndPuncttrue
\mciteSetBstMidEndSepPunct{\mcitedefaultmidpunct}
{\mcitedefaultendpunct}{\mcitedefaultseppunct}\relax
\EndOfBibitem
\bibitem[Marciniak \latin{et~al.}(2019)Marciniak, Despré, Loriot, Karras, Hervé, Quintard, Catoire, Joblin, Constant, Kuleff, and Lépine]{marciniak2019}
Marciniak,~A.; Despré,~V.; Loriot,~V.; Karras,~G.; Hervé,~M.; Quintard,~L.; Catoire,~F.; Joblin,~C.; Constant,~E.; Kuleff,~A.~I. \latin{et~al.}  Electron correlation driven non-adiabatic relaxation in molecules excited by an ultrashort extreme ultraviolet pulse. \emph{Nat. Commun.} \textbf{2019}, \emph{10}, 337\relax
\mciteBstWouldAddEndPuncttrue
\mciteSetBstMidEndSepPunct{\mcitedefaultmidpunct}
{\mcitedefaultendpunct}{\mcitedefaultseppunct}\relax
\EndOfBibitem
\bibitem[Monfared \latin{et~al.}(2022)Monfared, Irani, Lemell, and Burgd\"orfer]{monfared2022}
Monfared,~M.; Irani,~E.; Lemell,~C.; Burgd\"orfer,~J. Influence of coherent vibrational excitation on the high-order harmonic generation of diatomic molecules. \emph{Phys. Rev. A} \textbf{2022}, \emph{106}, 053108\relax
\mciteBstWouldAddEndPuncttrue
\mciteSetBstMidEndSepPunct{\mcitedefaultmidpunct}
{\mcitedefaultendpunct}{\mcitedefaultseppunct}\relax
\EndOfBibitem
\bibitem[McPherson \latin{et~al.}(1987)McPherson, Gibson, Jara, Johann, Luk, McIntyre, Boyer, and Rhodes]{mcpherson1987}
McPherson,~A.; Gibson,~G.; Jara,~H.; Johann,~U.; Luk,~T.~S.; McIntyre,~I.~A.; Boyer,~K.; Rhodes,~C.~K. Studies of multiphoton production of vacuum-ultraviolet radiation in the rare gases. \emph{J. Opt. Soc. Am. B} \textbf{1987}, \emph{4}, 595--601\relax
\mciteBstWouldAddEndPuncttrue
\mciteSetBstMidEndSepPunct{\mcitedefaultmidpunct}
{\mcitedefaultendpunct}{\mcitedefaultseppunct}\relax
\EndOfBibitem
\bibitem[Ferray \latin{et~al.}(1988)Ferray, L'Huillier, Li, Lompre, Mainfray, and Manus]{ferray1988}
Ferray,~M.; L'Huillier,~A.; Li,~X.~F.; Lompre,~L.~A.; Mainfray,~G.; Manus,~C. Multiple-harmonic conversion of 1064 nm radiation in rare gases. \emph{J. Phys. B: At. Mol. Opt. Phys.} \textbf{1988}, \emph{21}, L31\relax
\mciteBstWouldAddEndPuncttrue
\mciteSetBstMidEndSepPunct{\mcitedefaultmidpunct}
{\mcitedefaultendpunct}{\mcitedefaultseppunct}\relax
\EndOfBibitem
\bibitem[Li \latin{et~al.}(1989)Li, L'Huillier, Ferray, Lompr\'e, and Mainfray]{li1989}
Li,~X.~F.; L'Huillier,~A.; Ferray,~M.; Lompr\'e,~L.~A.; Mainfray,~G. Multiple-harmonic generation in rare gases at high laser intensity. \emph{Phys. Rev. A} \textbf{1989}, \emph{39}, 5751--5761\relax
\mciteBstWouldAddEndPuncttrue
\mciteSetBstMidEndSepPunct{\mcitedefaultmidpunct}
{\mcitedefaultendpunct}{\mcitedefaultseppunct}\relax
\EndOfBibitem
\bibitem[Sarukura \latin{et~al.}(1991)Sarukura, Hata, Adachi, Nodomi, Watanabe, and Watanabe]{sarukura1991}
Sarukura,~N.; Hata,~K.; Adachi,~T.; Nodomi,~R.; Watanabe,~M.; Watanabe,~S. Coherent soft-x-ray generation by the harmonics of an ultrahigh-power KrF laser. \emph{Phys. Rev. A} \textbf{1991}, \emph{43}, 1669--1672\relax
\mciteBstWouldAddEndPuncttrue
\mciteSetBstMidEndSepPunct{\mcitedefaultmidpunct}
{\mcitedefaultendpunct}{\mcitedefaultseppunct}\relax
\EndOfBibitem
\bibitem[Crane \latin{et~al.}(1992)Crane, Perry, Herman, and Falcone]{crane1992}
Crane,~J.~K.; Perry,~M.~D.; Herman,~S.; Falcone,~R.~W. High-field harmonic generation in helium. \emph{Opt. Lett.} \textbf{1992}, \emph{17}, 1256--1258\relax
\mciteBstWouldAddEndPuncttrue
\mciteSetBstMidEndSepPunct{\mcitedefaultmidpunct}
{\mcitedefaultendpunct}{\mcitedefaultseppunct}\relax
\EndOfBibitem
\bibitem[Faldon \latin{et~al.}(1992)Faldon, Hutchinson, Marangos, Muffett, Smith, Tisch, and Wahlstrom]{faldon1992}
Faldon,~M.~E.; Hutchinson,~M.; Marangos,~J.; Muffett,~J.; Smith,~R.; Tisch,~J.; Wahlstrom,~C. Studies of time-resolved harmonic generation in intense laser fields in xenon. \emph{J. Opt. Soc. Am. B} \textbf{1992}, \emph{9}, 2094--2099\relax
\mciteBstWouldAddEndPuncttrue
\mciteSetBstMidEndSepPunct{\mcitedefaultmidpunct}
{\mcitedefaultendpunct}{\mcitedefaultseppunct}\relax
\EndOfBibitem
\bibitem[Miyazaki and Sakai(1992)Miyazaki, and Sakai]{miyazaki1992}
Miyazaki,~K.; Sakai,~H. High-order harmonic generation in rare gases with intense subpicosecond dye laser pulses. \emph{J. Phys. B: At. Mol. Opt. Phys.} \textbf{1992}, \emph{25}, L83\relax
\mciteBstWouldAddEndPuncttrue
\mciteSetBstMidEndSepPunct{\mcitedefaultmidpunct}
{\mcitedefaultendpunct}{\mcitedefaultseppunct}\relax
\EndOfBibitem
\bibitem[L'Huillier and Balcou(1993)L'Huillier, and Balcou]{lhuillier1993}
L'Huillier,~A.; Balcou,~P. High-order harmonic generation in rare gases with a 1-ps 1053-nm laser. \emph{Phys. Rev. Lett.} \textbf{1993}, \emph{70}, 774--777\relax
\mciteBstWouldAddEndPuncttrue
\mciteSetBstMidEndSepPunct{\mcitedefaultmidpunct}
{\mcitedefaultendpunct}{\mcitedefaultseppunct}\relax
\EndOfBibitem
\bibitem[Kondo \latin{et~al.}(1993)Kondo, Sarukura, Sajiki, and Watanabe]{kondo1993}
Kondo,~K.; Sarukura,~N.; Sajiki,~K.; Watanabe,~S. High-order harmonic generation by ultrashort KrF and Ti:sapphire lasers. \emph{Phys. Rev. A} \textbf{1993}, \emph{47}, R2480--R2483\relax
\mciteBstWouldAddEndPuncttrue
\mciteSetBstMidEndSepPunct{\mcitedefaultmidpunct}
{\mcitedefaultendpunct}{\mcitedefaultseppunct}\relax
\EndOfBibitem
\bibitem[Macklin \latin{et~al.}(1993)Macklin, Kmetec, and Gordon]{macklin1993}
Macklin,~J.~J.; Kmetec,~J.~D.; Gordon,~C.~L. High-order harmonic generation using intense femtosecond pulses. \emph{Phys. Rev. Lett.} \textbf{1993}, \emph{70}, 766--769\relax
\mciteBstWouldAddEndPuncttrue
\mciteSetBstMidEndSepPunct{\mcitedefaultmidpunct}
{\mcitedefaultendpunct}{\mcitedefaultseppunct}\relax
\EndOfBibitem
\bibitem[Wahlstr\"om \latin{et~al.}(1993)Wahlstr\"om, Larsson, Persson, Starczewski, Svanberg, Sali\`eres, Balcou, and L'Huillier]{wahlstrom1993}
Wahlstr\"om,~C.-G.; Larsson,~J.; Persson,~A.; Starczewski,~T.; Svanberg,~S.; Sali\`eres,~P.; Balcou,~P.; L'Huillier,~A. High-order harmonic generation in rare gases with an intense short-pulse laser. \emph{Phys. Rev. A} \textbf{1993}, \emph{48}, 4709--4720\relax
\mciteBstWouldAddEndPuncttrue
\mciteSetBstMidEndSepPunct{\mcitedefaultmidpunct}
{\mcitedefaultendpunct}{\mcitedefaultseppunct}\relax
\EndOfBibitem
\bibitem[Liang \latin{et~al.}(1994)Liang, Augst, Chin, Beaudoin, and Chaker]{liang1994}
Liang,~Y.; Augst,~S.; Chin,~S.; Beaudoin,~Y.; Chaker,~M. High harmonic generation in atomic and diatomic molecular gases using intense picosecond laser pulses-a comparison. \emph{J. Phys. B: At. Mol. Opt. Phys.} \textbf{1994}, \emph{27}, 5119\relax
\mciteBstWouldAddEndPuncttrue
\mciteSetBstMidEndSepPunct{\mcitedefaultmidpunct}
{\mcitedefaultendpunct}{\mcitedefaultseppunct}\relax
\EndOfBibitem
\bibitem[Chin \latin{et~al.}(1995)Chin, Liang, Augst, Golovinski, Beaudoin, and Chaker]{chin1995}
Chin,~S.; Liang,~Y.; Augst,~S.; Golovinski,~P.; Beaudoin,~Y.; Chaker,~M. High Harmonic Generation in Atoms and Diatomic Molecules Using Ultrashort Laser Pulses in the Multiphoton Regime. \emph{J. Nonlinear Opt. Phys. Mater.} \textbf{1995}, \emph{4}, 667--686\relax
\mciteBstWouldAddEndPuncttrue
\mciteSetBstMidEndSepPunct{\mcitedefaultmidpunct}
{\mcitedefaultendpunct}{\mcitedefaultseppunct}\relax
\EndOfBibitem
\bibitem[Lyngå \latin{et~al.}(1996)Lyngå, L'Huillier, and Wahlström]{lynga1996}
Lyngå,~C.; L'Huillier,~A.; Wahlström,~C.-G. High-order harmonic generation in molecular gases. \emph{J. Phys. B: At. Mol. Opt. Phys.} \textbf{1996}, \emph{29}, 3293\relax
\mciteBstWouldAddEndPuncttrue
\mciteSetBstMidEndSepPunct{\mcitedefaultmidpunct}
{\mcitedefaultendpunct}{\mcitedefaultseppunct}\relax
\EndOfBibitem
\bibitem[Krause \latin{et~al.}(1992)Krause, Schafer, and Kulander]{krause1992a}
Krause,~J.~L.; Schafer,~K.~J.; Kulander,~K.~C. Calculation of photoemission from atoms subject to intense laser fields. \emph{Phys. Rev. A} \textbf{1992}, \emph{45}, 4998\relax
\mciteBstWouldAddEndPuncttrue
\mciteSetBstMidEndSepPunct{\mcitedefaultmidpunct}
{\mcitedefaultendpunct}{\mcitedefaultseppunct}\relax
\EndOfBibitem
\bibitem[Krause \latin{et~al.}(1992)Krause, Schafer, and Kulander]{krause1992b}
Krause,~J.~L.; Schafer,~K.~J.; Kulander,~K.~C. High-order harmonic generation from atoms and ions in the high intensity regime. \emph{Phys. Rev. Lett.} \textbf{1992}, \emph{68}, 3535--3538\relax
\mciteBstWouldAddEndPuncttrue
\mciteSetBstMidEndSepPunct{\mcitedefaultmidpunct}
{\mcitedefaultendpunct}{\mcitedefaultseppunct}\relax
\EndOfBibitem
\bibitem[Schafer \latin{et~al.}(1993)Schafer, Yang, DiMauro, and Kulander]{schafer1993}
Schafer,~K.~J.; Yang,~B.; DiMauro,~L.~F.; Kulander,~K.~C. Above threshold ionization beyond the high harmonic cutoff. \emph{Phys. Rev. Lett.} \textbf{1993}, \emph{70}, 1599\relax
\mciteBstWouldAddEndPuncttrue
\mciteSetBstMidEndSepPunct{\mcitedefaultmidpunct}
{\mcitedefaultendpunct}{\mcitedefaultseppunct}\relax
\EndOfBibitem
\bibitem[Corkum(1993)]{corkum1993}
Corkum,~P.~B. Plasma perspective on strong field multiphoton ionization. \emph{Phys. Rev. Lett.} \textbf{1993}, \emph{71}, 1994--1997\relax
\mciteBstWouldAddEndPuncttrue
\mciteSetBstMidEndSepPunct{\mcitedefaultmidpunct}
{\mcitedefaultendpunct}{\mcitedefaultseppunct}\relax
\EndOfBibitem
\bibitem[Lewenstein \latin{et~al.}(1994)Lewenstein, Balcou, Ivanov, L'Huillier, and Corkum]{lewenstein1994}
Lewenstein,~M.; Balcou,~P.; Ivanov,~M.~Y.; L'Huillier,~A.; Corkum,~P.~B. Theory of high-harmonic generation by low-frequency laser fields. \emph{Phys. Rev. A} \textbf{1994}, \emph{49}, 2117--2132\relax
\mciteBstWouldAddEndPuncttrue
\mciteSetBstMidEndSepPunct{\mcitedefaultmidpunct}
{\mcitedefaultendpunct}{\mcitedefaultseppunct}\relax
\EndOfBibitem
\bibitem[Constant \latin{et~al.}(1999)Constant, Garzella, Breger, M\'evel, Dorrer, Le~Blanc, Salin, and Agostini]{constant1999}
Constant,~E.; Garzella,~D.; Breger,~P.; M\'evel,~E.; Dorrer,~C.; Le~Blanc,~C.; Salin,~F.; Agostini,~P. Optimizing High Harmonic Generation in Absorbing Gases: Model and Experiment. \emph{Phys. Rev. Lett.} \textbf{1999}, \emph{82}, 1668--1671\relax
\mciteBstWouldAddEndPuncttrue
\mciteSetBstMidEndSepPunct{\mcitedefaultmidpunct}
{\mcitedefaultendpunct}{\mcitedefaultseppunct}\relax
\EndOfBibitem
\bibitem[Ghimire \latin{et~al.}(2011)Ghimire, DiChiara, Sistrunk, Agostini, DiMauro, and Reis]{ghimire2011}
Ghimire,~S.; DiChiara,~A.~D.; Sistrunk,~E.; Agostini,~P.; DiMauro,~L.~F.; Reis,~D.~A. Observation of high-order harmonic generation in a bulk crystal. \emph{Nat. Phys.} \textbf{2011}, \emph{7}, 138--141\relax
\mciteBstWouldAddEndPuncttrue
\mciteSetBstMidEndSepPunct{\mcitedefaultmidpunct}
{\mcitedefaultendpunct}{\mcitedefaultseppunct}\relax
\EndOfBibitem
\bibitem[Schubert \latin{et~al.}(2014)Schubert, Hohenleutner, Langer, Urbanek, Lange, Huttner, Golde, Meier, Kira, Koch, \latin{et~al.} others]{schubert2014}
Schubert,~O.; Hohenleutner,~M.; Langer,~F.; Urbanek,~B.; Lange,~C.; Huttner,~U.; Golde,~D.; Meier,~T.; Kira,~M.; Koch,~S.~W. \latin{et~al.}  Sub-cycle control of terahertz high-harmonic generation by dynamical Bloch oscillations. \emph{Nat. Photonics} \textbf{2014}, \emph{8}, 119--123\relax
\mciteBstWouldAddEndPuncttrue
\mciteSetBstMidEndSepPunct{\mcitedefaultmidpunct}
{\mcitedefaultendpunct}{\mcitedefaultseppunct}\relax
\EndOfBibitem
\bibitem[Luu \latin{et~al.}(2015)Luu, Garg, Kruchinin, Moulet, Hassan, and Goulielmakis]{luu2015}
Luu,~T.~T.; Garg,~M.; Kruchinin,~S.~Y.; Moulet,~A.; Hassan,~M.~T.; Goulielmakis,~E. Extreme ultraviolet high-harmonic spectroscopy of solids. \emph{Nature} \textbf{2015}, \emph{521}, 498--502\relax
\mciteBstWouldAddEndPuncttrue
\mciteSetBstMidEndSepPunct{\mcitedefaultmidpunct}
{\mcitedefaultendpunct}{\mcitedefaultseppunct}\relax
\EndOfBibitem
\bibitem[Ndabashimiye \latin{et~al.}(2016)Ndabashimiye, Ghimire, Wu, Browne, Schafer, Gaarde, and Reis]{ndabashimiye2016}
Ndabashimiye,~G.; Ghimire,~S.; Wu,~M.; Browne,~D.~A.; Schafer,~K.~J.; Gaarde,~M.~B.; Reis,~D.~A. Solid-state harmonics beyond the atomic limit. \emph{Nature} \textbf{2016}, \emph{534}, 520--523\relax
\mciteBstWouldAddEndPuncttrue
\mciteSetBstMidEndSepPunct{\mcitedefaultmidpunct}
{\mcitedefaultendpunct}{\mcitedefaultseppunct}\relax
\EndOfBibitem
\bibitem[You \latin{et~al.}(2017)You, Reis, and Ghimire]{you2017}
You,~Y.~S.; Reis,~D.~A.; Ghimire,~S. Anisotropic high-harmonic generation in bulk crystals. \emph{Nat. Phys.} \textbf{2017}, \emph{13}, 345--349\relax
\mciteBstWouldAddEndPuncttrue
\mciteSetBstMidEndSepPunct{\mcitedefaultmidpunct}
{\mcitedefaultendpunct}{\mcitedefaultseppunct}\relax
\EndOfBibitem
\bibitem[Luu \latin{et~al.}(2018)Luu, Yin, Jain, Gaumnitz, Pertot, Ma, and W{\"o}rner]{luu2018}
Luu,~T.~T.; Yin,~Z.; Jain,~A.; Gaumnitz,~T.; Pertot,~Y.; Ma,~J.; W{\"o}rner,~H.~J. Extreme--ultraviolet high--harmonic generation in liquids. \emph{Nat. Commun.} \textbf{2018}, \emph{9}, 3723\relax
\mciteBstWouldAddEndPuncttrue
\mciteSetBstMidEndSepPunct{\mcitedefaultmidpunct}
{\mcitedefaultendpunct}{\mcitedefaultseppunct}\relax
\EndOfBibitem
\bibitem[Golde \latin{et~al.}(2008)Golde, Meier, and Koch]{golde2008}
Golde,~D.; Meier,~T.; Koch,~S.~W. High harmonics generated in semiconductor nanostructures by the coupled dynamics of optical inter- and intraband excitations. \emph{Phys. Rev. B} \textbf{2008}, \emph{77}, 075330\relax
\mciteBstWouldAddEndPuncttrue
\mciteSetBstMidEndSepPunct{\mcitedefaultmidpunct}
{\mcitedefaultendpunct}{\mcitedefaultseppunct}\relax
\EndOfBibitem
\bibitem[Singhal \latin{et~al.}(2010)Singhal, Ganeev, Naik, Srivastava, Singh, Chari, Khan, Chakera, and Gupta]{singhal2010}
Singhal,~H.; Ganeev,~R.; Naik,~P.; Srivastava,~A.; Singh,~A.; Chari,~R.; Khan,~R.; Chakera,~J.; Gupta,~P. Study of high-order harmonic generation from nanoparticles. \emph{J. Phys. B: At. Mol. Opt. Phys.} \textbf{2010}, \emph{43}, 025603\relax
\mciteBstWouldAddEndPuncttrue
\mciteSetBstMidEndSepPunct{\mcitedefaultmidpunct}
{\mcitedefaultendpunct}{\mcitedefaultseppunct}\relax
\EndOfBibitem
\bibitem[Han \latin{et~al.}(2016)Han, Kim, Kim, Kim, Kim, Park, and Kim]{han2016}
Han,~S.; Kim,~H.; Kim,~Y.~W.; Kim,~Y.-J.; Kim,~S.; Park,~I.-Y.; Kim,~S.-W. High-harmonic generation by field enhanced femtosecond pulses in metal-sapphire nanostructure. \emph{Nat. Commun.} \textbf{2016}, \emph{7}, 13105\relax
\mciteBstWouldAddEndPuncttrue
\mciteSetBstMidEndSepPunct{\mcitedefaultmidpunct}
{\mcitedefaultendpunct}{\mcitedefaultseppunct}\relax
\EndOfBibitem
\bibitem[Liu \latin{et~al.}(2017)Liu, Li, You, Ghimire, Heinz, and Reis]{liu2017}
Liu,~H.; Li,~Y.; You,~Y.~S.; Ghimire,~S.; Heinz,~T.~F.; Reis,~D.~A. High-harmonic generation from an atomically thin semiconductor. \emph{Nat. Phys.} \textbf{2017}, \emph{13}, 262--265\relax
\mciteBstWouldAddEndPuncttrue
\mciteSetBstMidEndSepPunct{\mcitedefaultmidpunct}
{\mcitedefaultendpunct}{\mcitedefaultseppunct}\relax
\EndOfBibitem
\bibitem[Sivis \latin{et~al.}(2017)Sivis, Taucer, Vampa, Johnston, Staudte, Naumov, Villeneuve, Ropers, and Corkum]{sivis2017}
Sivis,~M.; Taucer,~M.; Vampa,~G.; Johnston,~K.; Staudte,~A.; Naumov,~A.~Y.; Villeneuve,~D.; Ropers,~C.; Corkum,~P. Tailored semiconductors for high-harmonic optoelectronics. \emph{Science} \textbf{2017}, \emph{357}, 303--306\relax
\mciteBstWouldAddEndPuncttrue
\mciteSetBstMidEndSepPunct{\mcitedefaultmidpunct}
{\mcitedefaultendpunct}{\mcitedefaultseppunct}\relax
\EndOfBibitem
\bibitem[Shcherbakov \latin{et~al.}(2021)Shcherbakov, Zhang, Tripepi, Sartorello, Talisa, AlShafey, Fan, Twardowski, Krivitsky, Kuznetsov, \latin{et~al.} others]{shcherbakov2021}
Shcherbakov,~M.~R.; Zhang,~H.; Tripepi,~M.; Sartorello,~G.; Talisa,~N.; AlShafey,~A.; Fan,~Z.; Twardowski,~J.; Krivitsky,~L.~A.; Kuznetsov,~A.~I. \latin{et~al.}  Generation of even and odd high harmonics in resonant metasurfaces using single and multiple ultra-intense laser pulses. \emph{Nat. Commun.} \textbf{2021}, \emph{12}, 4185\relax
\mciteBstWouldAddEndPuncttrue
\mciteSetBstMidEndSepPunct{\mcitedefaultmidpunct}
{\mcitedefaultendpunct}{\mcitedefaultseppunct}\relax
\EndOfBibitem
\bibitem[Lopata and Govind(2011)Lopata, and Govind]{lopata2011}
Lopata,~K.; Govind,~N. Modeling Fast Electron Dynamics with Real-Time Time-Dependent Density Functional Theory: Application to Small Molecules and Chromophores. \emph{J. Chem. Theory Comput.} \textbf{2011}, \emph{7}, 1344--1355\relax
\mciteBstWouldAddEndPuncttrue
\mciteSetBstMidEndSepPunct{\mcitedefaultmidpunct}
{\mcitedefaultendpunct}{\mcitedefaultseppunct}\relax
\EndOfBibitem
\bibitem[Ishikawa and Sato(2015)Ishikawa, and Sato]{ishikawa2015}
Ishikawa,~K.~L.; Sato,~T. A review on ab initio approaches for multielectron dynamics. \emph{IEEE J. Sel. Top. Quantum Electron.} \textbf{2015}, \emph{21}, 1--16\relax
\mciteBstWouldAddEndPuncttrue
\mciteSetBstMidEndSepPunct{\mcitedefaultmidpunct}
{\mcitedefaultendpunct}{\mcitedefaultseppunct}\relax
\EndOfBibitem
\bibitem[Goings \latin{et~al.}(2017)Goings, Lestrange, and Li]{goings2017}
Goings,~J.~J.; Lestrange,~P.~J.; Li,~X. Real‐time time‐dependent electronic structure theory. \emph{WIREs Comput. Mol. Sci.} \textbf{2017}, \emph{8}, e1341\relax
\mciteBstWouldAddEndPuncttrue
\mciteSetBstMidEndSepPunct{\mcitedefaultmidpunct}
{\mcitedefaultendpunct}{\mcitedefaultseppunct}\relax
\EndOfBibitem
\bibitem[Bedurke \latin{et~al.}(2021)Bedurke, Klamroth, and Saalfrank]{bedurke2021}
Bedurke,~F.; Klamroth,~T.; Saalfrank,~P. Many-electron dynamics in laser-driven molecules: wavefunction theory vs. density functional theory. \emph{Phys. Chem. Chem. Phys.} \textbf{2021}, \emph{23}, 13544--13560\relax
\mciteBstWouldAddEndPuncttrue
\mciteSetBstMidEndSepPunct{\mcitedefaultmidpunct}
{\mcitedefaultendpunct}{\mcitedefaultseppunct}\relax
\EndOfBibitem
\bibitem[Coccia and Luppi(2022)Coccia, and Luppi]{coccia2022}
Coccia,~E.; Luppi,~E. Time-dependent ab initio approaches for high-harmonic generation spectroscopy. \emph{J. Phys. Condens. Matter} \textbf{2022}, \emph{34}, 073001\relax
\mciteBstWouldAddEndPuncttrue
\mciteSetBstMidEndSepPunct{\mcitedefaultmidpunct}
{\mcitedefaultendpunct}{\mcitedefaultseppunct}\relax
\EndOfBibitem
\bibitem[Klamroth(2003)]{klamroth2003}
Klamroth,~T. Laser-driven electron transfer through metal-insulator-metal contacts: Time-dependent configuration interaction singles calculations for a jellium model. \emph{Phys. Rev. B} \textbf{2003}, \emph{68}, 245421\relax
\mciteBstWouldAddEndPuncttrue
\mciteSetBstMidEndSepPunct{\mcitedefaultmidpunct}
{\mcitedefaultendpunct}{\mcitedefaultseppunct}\relax
\EndOfBibitem
\bibitem[Krause \latin{et~al.}(2005)Krause, Klamroth, and Saalfrank]{krause2005}
Krause,~P.; Klamroth,~T.; Saalfrank,~P. Time-dependent configuration-interaction calculations of laser-pulse-driven many-electron dynamics: Controlled dipole switching in lithium cyanide. \emph{J. Chem. Phys.} \textbf{2005}, \emph{123}, 074105\relax
\mciteBstWouldAddEndPuncttrue
\mciteSetBstMidEndSepPunct{\mcitedefaultmidpunct}
{\mcitedefaultendpunct}{\mcitedefaultseppunct}\relax
\EndOfBibitem
\bibitem[Huber and Klamroth(2005)Huber, and Klamroth]{huber2005}
Huber,~C.; Klamroth,~T. Simulation of two-photon-photoelectron spectra at a jellium-vacuum interface. \emph{Appl. Phys. A} \textbf{2005}, \emph{81}, 93--101\relax
\mciteBstWouldAddEndPuncttrue
\mciteSetBstMidEndSepPunct{\mcitedefaultmidpunct}
{\mcitedefaultendpunct}{\mcitedefaultseppunct}\relax
\EndOfBibitem
\bibitem[Rohringer \latin{et~al.}(2006)Rohringer, Gordon, and Santra]{rohringer2006}
Rohringer,~N.; Gordon,~A.; Santra,~R. Configuration-interaction-based time-dependent orbital approach for ab initio treatment of electronic dynamics in a strong optical laser field. \emph{Phys. Rev. A} \textbf{2006}, \emph{74}, 043420\relax
\mciteBstWouldAddEndPuncttrue
\mciteSetBstMidEndSepPunct{\mcitedefaultmidpunct}
{\mcitedefaultendpunct}{\mcitedefaultseppunct}\relax
\EndOfBibitem
\bibitem[Krause \latin{et~al.}(2007)Krause, Klamroth, and Saalfrank]{krause2007}
Krause,~P.; Klamroth,~T.; Saalfrank,~P. Molecular response properties from explicitly time-dependent configuration interaction methods. \emph{J. Chem. Phys.} \textbf{2007}, \emph{127}, 034107\relax
\mciteBstWouldAddEndPuncttrue
\mciteSetBstMidEndSepPunct{\mcitedefaultmidpunct}
{\mcitedefaultendpunct}{\mcitedefaultseppunct}\relax
\EndOfBibitem
\bibitem[Schlegel \latin{et~al.}(2007)Schlegel, Smith, and Li]{schlegel2007}
Schlegel,~H.~B.; Smith,~S.~M.; Li,~X. Electronic optical response of molecules in intense fields: Comparison of TD-HF, TD-CIS, and TD-CIS(D) approaches. \emph{J. Chem. Phys.} \textbf{2007}, \emph{126}, 244110\relax
\mciteBstWouldAddEndPuncttrue
\mciteSetBstMidEndSepPunct{\mcitedefaultmidpunct}
{\mcitedefaultendpunct}{\mcitedefaultseppunct}\relax
\EndOfBibitem
\bibitem[Luppi and Head-Gordon(2013)Luppi, and Head-Gordon]{luppi2013}
Luppi,~E.; Head-Gordon,~M. The role of Rydberg and continuum levels in computing high harmonic generation spectra of the hydrogen atom using time-dependent configuration interaction. \emph{J. Chem. Phys.} \textbf{2013}, \emph{139}, 164121\relax
\mciteBstWouldAddEndPuncttrue
\mciteSetBstMidEndSepPunct{\mcitedefaultmidpunct}
{\mcitedefaultendpunct}{\mcitedefaultseppunct}\relax
\EndOfBibitem
\bibitem[Coccia \latin{et~al.}(2016)Coccia, Mussard, Labeye, Caillat, Ta{\"\i}eb, Toulouse, and Luppi]{coccia2016a}
Coccia,~E.; Mussard,~B.; Labeye,~M.; Caillat,~J.; Ta{\"\i}eb,~R.; Toulouse,~J.; Luppi,~E. Gaussian continuum basis functions for calculating high-harmonic generation spectra. \emph{Int. J. Quantum Chem.} \textbf{2016}, \emph{116}, 1120--1131\relax
\mciteBstWouldAddEndPuncttrue
\mciteSetBstMidEndSepPunct{\mcitedefaultmidpunct}
{\mcitedefaultendpunct}{\mcitedefaultseppunct}\relax
\EndOfBibitem
\bibitem[Coccia and Luppi(2016)Coccia, and Luppi]{coccia2016b}
Coccia,~E.; Luppi,~E. Optimal-continuum and multicentered Gaussian basis sets for high-harmonic generation spectroscopy. \emph{Theor. Chem. Acc.} \textbf{2016}, \emph{135}, 43\relax
\mciteBstWouldAddEndPuncttrue
\mciteSetBstMidEndSepPunct{\mcitedefaultmidpunct}
{\mcitedefaultendpunct}{\mcitedefaultseppunct}\relax
\EndOfBibitem
\bibitem[Pabst \latin{et~al.}(2016)Pabst, Sytcheva, Geffert, and Santra]{pabst2016}
Pabst,~S.; Sytcheva,~A.; Geffert,~O.; Santra,~R. Stability of the time-dependent configuration-interaction-singles method in the attosecond and strong-field regimes: A study of basis sets and absorption methods. \emph{Phys. Rev. A} \textbf{2016}, \emph{94}, 033421\relax
\mciteBstWouldAddEndPuncttrue
\mciteSetBstMidEndSepPunct{\mcitedefaultmidpunct}
{\mcitedefaultendpunct}{\mcitedefaultseppunct}\relax
\EndOfBibitem
\bibitem[Coccia and Luppi(2019)Coccia, and Luppi]{coccia2019}
Coccia,~E.; Luppi,~E. Detecting the minimum in argon high-harmonic generation spectrum using Gaussian basis sets. \emph{Theor. Chem. Acc.} \textbf{2019}, \emph{138}, 96\relax
\mciteBstWouldAddEndPuncttrue
\mciteSetBstMidEndSepPunct{\mcitedefaultmidpunct}
{\mcitedefaultendpunct}{\mcitedefaultseppunct}\relax
\EndOfBibitem
\bibitem[Coccia(2020)]{coccia2020}
Coccia,~E. How electronic dephasing affects the high-harmonic generation in atoms. \emph{Mol. Phys.} \textbf{2020}, \emph{118}, e1769871\relax
\mciteBstWouldAddEndPuncttrue
\mciteSetBstMidEndSepPunct{\mcitedefaultmidpunct}
{\mcitedefaultendpunct}{\mcitedefaultseppunct}\relax
\EndOfBibitem
\bibitem[Wo{\'z}niak \latin{et~al.}(2021)Wo{\'z}niak, Lesiuk, Przybytek, Efimov, Prauzner-Bechcicki, Mandrysz, Ciappina, Pisanty, Zakrzewski, Lewenstein, \latin{et~al.} others]{wozniak2021}
Wo{\'z}niak,~A.~P.; Lesiuk,~M.; Przybytek,~M.; Efimov,~D.~K.; Prauzner-Bechcicki,~J.~S.; Mandrysz,~M.; Ciappina,~M.; Pisanty,~E.; Zakrzewski,~J.; Lewenstein,~M. \latin{et~al.}  A systematic construction of Gaussian basis sets for the description of laser field ionization and high-harmonic generation. \emph{J. Chem. Phys.} \textbf{2021}, \emph{154}, 094111\relax
\mciteBstWouldAddEndPuncttrue
\mciteSetBstMidEndSepPunct{\mcitedefaultmidpunct}
{\mcitedefaultendpunct}{\mcitedefaultseppunct}\relax
\EndOfBibitem
\bibitem[Wo{\'z}niak \latin{et~al.}(2022)Wo{\'z}niak, Przybytek, Lewenstein, and Moszyński]{wozniak2022}
Wo{\'z}niak,~A.~P.; Przybytek,~M.; Lewenstein,~M.; Moszyński,~R. Effects of electronic correlation on the high harmonic generation in helium: A time-dependent configuration interaction singles vs time-dependent full configuration interaction study. \emph{J. Chem. Phys.} \textbf{2022}, \emph{156}, 174106\relax
\mciteBstWouldAddEndPuncttrue
\mciteSetBstMidEndSepPunct{\mcitedefaultmidpunct}
{\mcitedefaultendpunct}{\mcitedefaultseppunct}\relax
\EndOfBibitem
\bibitem[Luppi and Head-Gordon(2012)Luppi, and Head-Gordon]{luppi2012}
Luppi,~E.; Head-Gordon,~M. Computation of high-harmonic generation spectra of H2 And N2 in intense laser pulses using quantum chemistry methods and time-dependent density functional theory. \emph{Mol. Phys.} \textbf{2012}, \emph{110}, 909--923\relax
\mciteBstWouldAddEndPuncttrue
\mciteSetBstMidEndSepPunct{\mcitedefaultmidpunct}
{\mcitedefaultendpunct}{\mcitedefaultseppunct}\relax
\EndOfBibitem
\bibitem[White \latin{et~al.}(2016)White, Heide, Saalfrank, Head-Gordon, and Luppi]{white2016}
White,~A.~F.; Heide,~C.~J.; Saalfrank,~P.; Head-Gordon,~M.; Luppi,~E. Computation of high-harmonic generation spectra of the hydrogen molecule using time-dependent configuration-interaction. \emph{Mol. Phys.} \textbf{2016}, \emph{114}, 947--956\relax
\mciteBstWouldAddEndPuncttrue
\mciteSetBstMidEndSepPunct{\mcitedefaultmidpunct}
{\mcitedefaultendpunct}{\mcitedefaultseppunct}\relax
\EndOfBibitem
\bibitem[Labeye \latin{et~al.}(2018)Labeye, Zapata, Coccia, V{\'e}niard, Toulouse, Caillat, Ta{\"\i}eb, and Luppi]{labeye2018}
Labeye,~M.; Zapata,~F.; Coccia,~E.; V{\'e}niard,~V.; Toulouse,~J.; Caillat,~J.; Ta{\"\i}eb,~R.; Luppi,~E. Optimal Basis Set for Electron Dynamics in Strong Laser Fields: The case of Molecular Ion H2+. \emph{J. Chem. Theory Comput.} \textbf{2018}, \emph{14}, 5846--5858\relax
\mciteBstWouldAddEndPuncttrue
\mciteSetBstMidEndSepPunct{\mcitedefaultmidpunct}
{\mcitedefaultendpunct}{\mcitedefaultseppunct}\relax
\EndOfBibitem
\bibitem[Pauletti \latin{et~al.}(2021)Pauletti, Coccia, and Luppi]{pauletti2021}
Pauletti,~C.~F.; Coccia,~E.; Luppi,~E. Role of exchange and correlation in high-harmonic generation spectra of H2, N2, and CO2: Real-time time-dependent electronic-structure approaches. \emph{J. Chem. Phys.} \textbf{2021}, \emph{154}, 014101\relax
\mciteBstWouldAddEndPuncttrue
\mciteSetBstMidEndSepPunct{\mcitedefaultmidpunct}
{\mcitedefaultendpunct}{\mcitedefaultseppunct}\relax
\EndOfBibitem
\bibitem[Wo{\'z}niak \latin{et~al.}(2023)Wo{\'z}niak, Lewenstein, and Moszyński]{wozniak2023}
Wo{\'z}niak,~A.~P.; Lewenstein,~M.; Moszyński,~R. In \emph{Polish Quantum Chemistry from Kołos to Now}; Musiał,~M., Grabowski,~I., Eds.; Advances in Quantum Chemistry; Academic Press, 2023; Vol.~87; pp 167--190\relax
\mciteBstWouldAddEndPuncttrue
\mciteSetBstMidEndSepPunct{\mcitedefaultmidpunct}
{\mcitedefaultendpunct}{\mcitedefaultseppunct}\relax
\EndOfBibitem
\bibitem[Bedurke \latin{et~al.}(2019)Bedurke, Klamroth, Krause, and Saalfrank]{bedurke2019}
Bedurke,~F.; Klamroth,~T.; Krause,~P.; Saalfrank,~P. Discriminating organic isomers by high harmonic generation: A time-dependent configuration interaction singles study. \emph{J. Chem. Phys.} \textbf{2019}, \emph{150}, 234114\relax
\mciteBstWouldAddEndPuncttrue
\mciteSetBstMidEndSepPunct{\mcitedefaultmidpunct}
{\mcitedefaultendpunct}{\mcitedefaultseppunct}\relax
\EndOfBibitem
\bibitem[Luppi and Coccia(2021)Luppi, and Coccia]{luppi2021}
Luppi,~E.; Coccia,~E. Probing the molecular frame of uracil and thymine with high-harmonic generation spectroscopy. \emph{Phys. Chem. Chem. Phys.} \textbf{2021}, \emph{23}, 3729--3738\relax
\mciteBstWouldAddEndPuncttrue
\mciteSetBstMidEndSepPunct{\mcitedefaultmidpunct}
{\mcitedefaultendpunct}{\mcitedefaultseppunct}\relax
\EndOfBibitem
\bibitem[Morassut \latin{et~al.}(2022)Morassut, Luppi, and Coccia]{morassut2022}
Morassut,~C.; Luppi,~E.; Coccia,~E. A TD-CIS study of high-harmonic generation of uracil cation fragments. \emph{Chem. Phys.} \textbf{2022}, \emph{559}, 111515\relax
\mciteBstWouldAddEndPuncttrue
\mciteSetBstMidEndSepPunct{\mcitedefaultmidpunct}
{\mcitedefaultendpunct}{\mcitedefaultseppunct}\relax
\EndOfBibitem
\bibitem[Luppi and Coccia(2023)Luppi, and Coccia]{luppi2023}
Luppi,~E.; Coccia,~E. Role of Inner Molecular Orbitals in High-Harmonic Generation Spectra of Aligned Uracil. \emph{J. Phys. Chem. A} \textbf{2023}, \emph{127}, 7335--7343\relax
\mciteBstWouldAddEndPuncttrue
\mciteSetBstMidEndSepPunct{\mcitedefaultmidpunct}
{\mcitedefaultendpunct}{\mcitedefaultseppunct}\relax
\EndOfBibitem
\bibitem[Wo{\'z}niak \latin{et~al.}()Wo{\'z}niak, Lewenstein, and Moszyński]{wozniak2023b}
Wo{\'z}niak,~A.~P.; Lewenstein,~M.; Moszyński,~R. submitted to Phys. Rev. A\relax
\mciteBstWouldAddEndPuncttrue
\mciteSetBstMidEndSepPunct{\mcitedefaultmidpunct}
{\mcitedefaultendpunct}{\mcitedefaultseppunct}\relax
\EndOfBibitem
\bibitem[Ganeev \latin{et~al.}(2009)Ganeev, Bom, Abdul-Hadi, Wong, Brichta, Bhardwaj, and Ozaki]{ganeev2009a}
Ganeev,~R.~A.; Bom,~L. B.~E.; Abdul-Hadi,~J.; Wong,~M. C.~H.; Brichta,~J.~P.; Bhardwaj,~V.~R.; Ozaki,~T. Higher-Order Harmonic Generation from Fullerene by Means of the Plasma Harmonic Method. \emph{Phys. Rev. Lett.} \textbf{2009}, \emph{102}, 013903\relax
\mciteBstWouldAddEndPuncttrue
\mciteSetBstMidEndSepPunct{\mcitedefaultmidpunct}
{\mcitedefaultendpunct}{\mcitedefaultseppunct}\relax
\EndOfBibitem
\bibitem[Ganeev \latin{et~al.}(2009)Ganeev, Elouga~Bom, Wong, Brichta, Bhardwaj, Redkin, and Ozaki]{ganeev2009b}
Ganeev,~R.~A.; Elouga~Bom,~L.~B.; Wong,~M. C.~H.; Brichta,~J.-P.; Bhardwaj,~V.~R.; Redkin,~P.~V.; Ozaki,~T. High-order harmonic generation from ${\text{C}}_{60}$-rich plasma. \emph{Phys. Rev. A} \textbf{2009}, \emph{80}, 043808\relax
\mciteBstWouldAddEndPuncttrue
\mciteSetBstMidEndSepPunct{\mcitedefaultmidpunct}
{\mcitedefaultendpunct}{\mcitedefaultseppunct}\relax
\EndOfBibitem
\bibitem[Ganeev \latin{et~al.}(2009)Ganeev, Singhal, Naik, Chakera, Srivastava, Dhami, Joshi, and Gupta]{ganeev2009c}
Ganeev,~R.; Singhal,~H.; Naik,~P.; Chakera,~J.; Srivastava,~A.; Dhami,~T.; Joshi,~M.; Gupta,~P. Influence of C60 morphology on high-order harmonic generation enhancement in fullerene-containing plasma. \emph{J. Appl. Phys.} \textbf{2009}, \emph{106}\relax
\mciteBstWouldAddEndPuncttrue
\mciteSetBstMidEndSepPunct{\mcitedefaultmidpunct}
{\mcitedefaultendpunct}{\mcitedefaultseppunct}\relax
\EndOfBibitem
\bibitem[Ganeev \latin{et~al.}(2013)Ganeev, Hutchison, Witting, Frank, Weber, Okell, Fiordilino, Cricchio, Persico, Za{\"\i}r, \latin{et~al.} others]{ganeev2013}
Ganeev,~R.~A.; Hutchison,~C.; Witting,~T.; Frank,~F.; Weber,~S.; Okell,~W.~A.; Fiordilino,~E.; Cricchio,~D.; Persico,~F.; Za{\"\i}r,~A. \latin{et~al.}  High-order harmonic generation in fullerenes using few-and multi-cycle pulses of different wavelengths. \emph{J. Opt. Soc. Am. B} \textbf{2013}, \emph{30}, 7--12\relax
\mciteBstWouldAddEndPuncttrue
\mciteSetBstMidEndSepPunct{\mcitedefaultmidpunct}
{\mcitedefaultendpunct}{\mcitedefaultseppunct}\relax
\EndOfBibitem
\bibitem[Ganeev \latin{et~al.}(2016)Ganeev, Suzuki, and Kuroda]{ganeev2016}
Ganeev,~R.~A.; Suzuki,~M.; Kuroda,~H. High-order harmonic generation in Ag, Sn, fullerene, and graphene nanoparticle-contained plasmas using two-color mid-infrared pulses. \emph{Eur. Phys. J. D} \textbf{2016}, \emph{70}, 1--8\relax
\mciteBstWouldAddEndPuncttrue
\mciteSetBstMidEndSepPunct{\mcitedefaultmidpunct}
{\mcitedefaultendpunct}{\mcitedefaultseppunct}\relax
\EndOfBibitem
\bibitem[Ganeev(2011)]{ganeev2011}
Ganeev,~R. Fullerenes: the attractive medium for harmonic generation. \emph{Laser Phys.} \textbf{2011}, \emph{21}, 25--43\relax
\mciteBstWouldAddEndPuncttrue
\mciteSetBstMidEndSepPunct{\mcitedefaultmidpunct}
{\mcitedefaultendpunct}{\mcitedefaultseppunct}\relax
\EndOfBibitem
\bibitem[Ciappina \latin{et~al.}(2007)Ciappina, Becker, and Jaro\ifmmode \acute{n}\else~\'{n}\fi{} Becker]{ciappina2007}
Ciappina,~M.~F.; Becker,~A.; Jaro\ifmmode \acute{n}\else~\'{n}\fi{} Becker,~A. Multislit interference patterns in high-order harmonic generation in ${\mathrm{C}}_{60}$. \emph{Phys. Rev. A} \textbf{2007}, \emph{76}, 063406\relax
\mciteBstWouldAddEndPuncttrue
\mciteSetBstMidEndSepPunct{\mcitedefaultmidpunct}
{\mcitedefaultendpunct}{\mcitedefaultseppunct}\relax
\EndOfBibitem
\bibitem[Ciappina \latin{et~al.}(2008)Ciappina, Becker, and Jaro\ifmmode \acute{n}\else~\'{n}\fi{} Becker]{ciappina2008}
Ciappina,~M.~F.; Becker,~A.; Jaro\ifmmode \acute{n}\else~\'{n}\fi{} Becker,~A. High-order harmonic generation in fullerenes with icosahedral symmetry. \emph{Phys. Rev. A} \textbf{2008}, \emph{78}, 063405\relax
\mciteBstWouldAddEndPuncttrue
\mciteSetBstMidEndSepPunct{\mcitedefaultmidpunct}
{\mcitedefaultendpunct}{\mcitedefaultseppunct}\relax
\EndOfBibitem
\bibitem[Topcu \latin{et~al.}(2019)Topcu, Bleda, and Altun]{topcu2019}
Topcu,~T.; Bleda,~E.~A.; Altun,~Z. Drastically enhanced high-order harmonic generation from endofullerenes. \emph{Phys. Rev. A} \textbf{2019}, \emph{100}, 063421\relax
\mciteBstWouldAddEndPuncttrue
\mciteSetBstMidEndSepPunct{\mcitedefaultmidpunct}
{\mcitedefaultendpunct}{\mcitedefaultseppunct}\relax
\EndOfBibitem
\bibitem[Zhang(2005)]{zhang2005}
Zhang,~G.~P. Optical High Harmonic Generation in ${\mathrm{C}}_{60}$. \emph{Phys. Rev. Lett.} \textbf{2005}, \emph{95}, 047401\relax
\mciteBstWouldAddEndPuncttrue
\mciteSetBstMidEndSepPunct{\mcitedefaultmidpunct}
{\mcitedefaultendpunct}{\mcitedefaultseppunct}\relax
\EndOfBibitem
\bibitem[Zhang and George(2006)Zhang, and George]{zhang2006}
Zhang,~G.~P.; George,~T.~F. Ellipticity dependence of optical harmonic generation in ${\mathrm{C}}_{60}$. \emph{Phys. Rev. A} \textbf{2006}, \emph{74}, 023811\relax
\mciteBstWouldAddEndPuncttrue
\mciteSetBstMidEndSepPunct{\mcitedefaultmidpunct}
{\mcitedefaultendpunct}{\mcitedefaultseppunct}\relax
\EndOfBibitem
\bibitem[Avetissian \latin{et~al.}(2021)Avetissian, Ghazaryan, and Mkrtchian]{avetissian2021}
Avetissian,~H.~K.; Ghazaryan,~A.~G.; Mkrtchian,~G.~F. High harmonic generation in fullerene molecules. \emph{Phys. Rev. B} \textbf{2021}, \emph{104}, 125436\relax
\mciteBstWouldAddEndPuncttrue
\mciteSetBstMidEndSepPunct{\mcitedefaultmidpunct}
{\mcitedefaultendpunct}{\mcitedefaultseppunct}\relax
\EndOfBibitem
\bibitem[Redkin and Ganeev(2010)Redkin, and Ganeev]{redkin2010}
Redkin,~P.~V.; Ganeev,~R.~A. Simulation of resonant high-order harmonic generation in a three-dimensional fullerenelike system by means of a multiconfigurational time-dependent Hartree-Fock approach. \emph{Phys. Rev. A} \textbf{2010}, \emph{81}, 063825\relax
\mciteBstWouldAddEndPuncttrue
\mciteSetBstMidEndSepPunct{\mcitedefaultmidpunct}
{\mcitedefaultendpunct}{\mcitedefaultseppunct}\relax
\EndOfBibitem
\bibitem[Redkin \latin{et~al.}(2011)Redkin, Danailov, and Ganeev]{redkin2011}
Redkin,~P.~V.; Danailov,~M.~B.; Ganeev,~R.~A. Endohedral fullerenes: A way to control resonant high-order harmonic generation. \emph{Phys. Rev. A} \textbf{2011}, \emph{84}, 013407\relax
\mciteBstWouldAddEndPuncttrue
\mciteSetBstMidEndSepPunct{\mcitedefaultmidpunct}
{\mcitedefaultendpunct}{\mcitedefaultseppunct}\relax
\EndOfBibitem
\bibitem[Zhang and Bai(2020)Zhang, and Bai]{zhang2020}
Zhang,~G.~P.; Bai,~Y.~H. High-order harmonic generation in solid ${\mathrm{C}}_{60}$. \emph{Phys. Rev. B} \textbf{2020}, \emph{101}, 081412\relax
\mciteBstWouldAddEndPuncttrue
\mciteSetBstMidEndSepPunct{\mcitedefaultmidpunct}
{\mcitedefaultendpunct}{\mcitedefaultseppunct}\relax
\EndOfBibitem
\bibitem[Castro \latin{et~al.}(2015)Castro, Rubio, and Gross]{castro2015}
Castro,~A.; Rubio,~A.; Gross,~E. K.~U. Enhancing and controlling single-atom high-harmonic generation spectra: a time-dependent density-functional scheme. \emph{Eur. Phys. J. B} \textbf{2015}, \emph{88}, 191\relax
\mciteBstWouldAddEndPuncttrue
\mciteSetBstMidEndSepPunct{\mcitedefaultmidpunct}
{\mcitedefaultendpunct}{\mcitedefaultseppunct}\relax
\EndOfBibitem
\bibitem[Artemyev \latin{et~al.}(2017)Artemyev, Cederbaum, and Demekhin]{artemyev2017}
Artemyev,~A.~N.; Cederbaum,~L.~S.; Demekhin,~P.~V. Impact of two-electron dynamics and correlations on high-order-harmonic generation in He. \emph{Phys. Rev. A} \textbf{2017}, \emph{95}, 033402\relax
\mciteBstWouldAddEndPuncttrue
\mciteSetBstMidEndSepPunct{\mcitedefaultmidpunct}
{\mcitedefaultendpunct}{\mcitedefaultseppunct}\relax
\EndOfBibitem
\bibitem[Sato \latin{et~al.}(2018)Sato, Pathak, Orimo, and Ishikawa]{sato2018}
Sato,~T.; Pathak,~H.; Orimo,~Y.; Ishikawa,~K.~L. Communication: Time-dependent optimized coupled-cluster method for multielectron dynamics. \emph{J. Chem. Phys.} \textbf{2018}, \emph{148}, 051101\relax
\mciteBstWouldAddEndPuncttrue
\mciteSetBstMidEndSepPunct{\mcitedefaultmidpunct}
{\mcitedefaultendpunct}{\mcitedefaultseppunct}\relax
\EndOfBibitem
\bibitem[Neufeld and Cohen(2020)Neufeld, and Cohen]{neufeld2020}
Neufeld,~O.; Cohen,~O. Probing ultrafast electron correlations in high harmonic generation. \emph{Phys. Rev. Res.} \textbf{2020}, \emph{2}, 033037\relax
\mciteBstWouldAddEndPuncttrue
\mciteSetBstMidEndSepPunct{\mcitedefaultmidpunct}
{\mcitedefaultendpunct}{\mcitedefaultseppunct}\relax
\EndOfBibitem
\bibitem[Reiff \latin{et~al.}(2020)Reiff, Joyce, Jaro{\'n}-Becker, and Becker]{reiff2020}
Reiff,~R.; Joyce,~T.; Jaro{\'n}-Becker,~A.; Becker,~A. Single-active electron calculations of high-order harmonic generation from valence shells in atoms for quantitative comparison with TDDFT calculations. \emph{J. Phys. Commun.} \textbf{2020}, \emph{4}, 065011\relax
\mciteBstWouldAddEndPuncttrue
\mciteSetBstMidEndSepPunct{\mcitedefaultmidpunct}
{\mcitedefaultendpunct}{\mcitedefaultseppunct}\relax
\EndOfBibitem
\bibitem[Nguyen and Bandrauk(2006)Nguyen, and Bandrauk]{nguyen2006}
Nguyen,~N.~A.; Bandrauk,~A.~D. Electron correlation of one-dimensional ${\mathrm{H}}_{2}$ in intense laser fields: Time-dependent extended Hartree-Fock and time-dependent density-functional-theory approaches. \emph{Phys. Rev. A} \textbf{2006}, \emph{73}, 032708\relax
\mciteBstWouldAddEndPuncttrue
\mciteSetBstMidEndSepPunct{\mcitedefaultmidpunct}
{\mcitedefaultendpunct}{\mcitedefaultseppunct}\relax
\EndOfBibitem
\bibitem[Saalfrank \latin{et~al.}(2020)Saalfrank, Bedurke, Heide, Klamroth, Klinkusch, Krause, Nest, and Tremblay]{saalfrank2020}
Saalfrank,~P.; Bedurke,~F.; Heide,~C.; Klamroth,~T.; Klinkusch,~S.; Krause,~P.; Nest,~M.; Tremblay,~J.~C. In \emph{Chemical Physics and Quantum Chemistry}; Ruud,~K., Brändas,~E.~J., Eds.; Advances in Quantum Chemistry; Academic Press, 2020; Vol.~81; pp 15--50\relax
\mciteBstWouldAddEndPuncttrue
\mciteSetBstMidEndSepPunct{\mcitedefaultmidpunct}
{\mcitedefaultendpunct}{\mcitedefaultseppunct}\relax
\EndOfBibitem
\bibitem[St{\"u}ck \latin{et~al.}(2011)St{\"u}ck, Baker, Zimmerman, Kurlancheek, and Head-Gordon]{stuck2011}
St{\"u}ck,~D.; Baker,~T.~A.; Zimmerman,~P.; Kurlancheek,~W.; Head-Gordon,~M. On the nature of electron correlation in C60. \emph{J. Chem. Phys.} \textbf{2011}, \emph{135}, 11B608\relax
\mciteBstWouldAddEndPuncttrue
\mciteSetBstMidEndSepPunct{\mcitedefaultmidpunct}
{\mcitedefaultendpunct}{\mcitedefaultseppunct}\relax
\EndOfBibitem
\bibitem[Runge and Gross(1984)Runge, and Gross]{runge1984}
Runge,~E.; Gross,~E. K.~U. Density-functional theory for time-dependent systems. \emph{Phys. Rev. Lett.} \textbf{1984}, \emph{52}, 997--1000\relax
\mciteBstWouldAddEndPuncttrue
\mciteSetBstMidEndSepPunct{\mcitedefaultmidpunct}
{\mcitedefaultendpunct}{\mcitedefaultseppunct}\relax
\EndOfBibitem
\bibitem[Tong and Chu(1998)Tong, and Chu]{tong1998}
Tong,~X.-M.; Chu,~S.-I. Time-dependent density-functional theory for strong-field multiphoton processes: Application to the study of the role of dynamical electron correlation in multiple high-order harmonic generation. \emph{Phys. Rev. A} \textbf{1998}, \emph{57}, 452--461\relax
\mciteBstWouldAddEndPuncttrue
\mciteSetBstMidEndSepPunct{\mcitedefaultmidpunct}
{\mcitedefaultendpunct}{\mcitedefaultseppunct}\relax
\EndOfBibitem
\bibitem[Castro \latin{et~al.}(2004)Castro, Marques, and Rubio]{castro2004}
Castro,~A.; Marques,~M. A.~L.; Rubio,~A. Propagators for the time-dependent Kohn–Sham equations. \emph{J. Chem. Phys.} \textbf{2004}, \emph{121}, 3425--3433\relax
\mciteBstWouldAddEndPuncttrue
\mciteSetBstMidEndSepPunct{\mcitedefaultmidpunct}
{\mcitedefaultendpunct}{\mcitedefaultseppunct}\relax
\EndOfBibitem
\bibitem[Isborn and Li(2008)Isborn, and Li]{isborn2008}
Isborn,~C.~M.; Li,~X. Modeling the doubly excited state with time-dependent Hartree--Fock and density functional theories. \emph{J. Chem. Phys.} \textbf{2008}, \emph{129}\relax
\mciteBstWouldAddEndPuncttrue
\mciteSetBstMidEndSepPunct{\mcitedefaultmidpunct}
{\mcitedefaultendpunct}{\mcitedefaultseppunct}\relax
\EndOfBibitem
\bibitem[Habenicht \latin{et~al.}(2014)Habenicht, Tani, Provorse, and Isborn]{habenicht2014}
Habenicht,~B.~F.; Tani,~N.~P.; Provorse,~M.~R.; Isborn,~C.~M. Two-electron Rabi oscillations in real-time time-dependent density-functional theory. \emph{J. Chem. Phys.} \textbf{2014}, \emph{141}\relax
\mciteBstWouldAddEndPuncttrue
\mciteSetBstMidEndSepPunct{\mcitedefaultmidpunct}
{\mcitedefaultendpunct}{\mcitedefaultseppunct}\relax
\EndOfBibitem
\bibitem[Castro \latin{et~al.}(2012)Castro, Werschnik, and Gross]{castro2012}
Castro,~A.; Werschnik,~J.; Gross,~E. K.~U. Controlling the Dynamics of Many-Electron Systems from First Principles: A Combination of Optimal Control and Time-Dependent Density-Functional Theory. \emph{Phys. Rev. Lett.} \textbf{2012}, \emph{109}, 153603\relax
\mciteBstWouldAddEndPuncttrue
\mciteSetBstMidEndSepPunct{\mcitedefaultmidpunct}
{\mcitedefaultendpunct}{\mcitedefaultseppunct}\relax
\EndOfBibitem
\bibitem[Kvaal(2012)]{kvaal2012}
Kvaal,~S. Ab initio quantum dynamics using coupled-cluster. \emph{J. Chem. Phys.} \textbf{2012}, \emph{136}, 194109\relax
\mciteBstWouldAddEndPuncttrue
\mciteSetBstMidEndSepPunct{\mcitedefaultmidpunct}
{\mcitedefaultendpunct}{\mcitedefaultseppunct}\relax
\EndOfBibitem
\bibitem[Ridley and Zerner(1973)Ridley, and Zerner]{ridley1973}
Ridley,~J.; Zerner,~M. An intermediate neglect of differential overlap technique for spectroscopy: pyrrole and the azines. \emph{Theor. Chim. Acta} \textbf{1973}, \emph{32}, 111--134\relax
\mciteBstWouldAddEndPuncttrue
\mciteSetBstMidEndSepPunct{\mcitedefaultmidpunct}
{\mcitedefaultendpunct}{\mcitedefaultseppunct}\relax
\EndOfBibitem
\bibitem[Ghosh \latin{et~al.}(2017)Ghosh, Andersen, Gagliardi, Cramer, and Govind]{ghosh2017}
Ghosh,~S.; Andersen,~A.; Gagliardi,~L.; Cramer,~C.~J.; Govind,~N. Modeling optical spectra of large organic systems using real-time propagation of semiempirical effective Hamiltonians. \emph{J. Chem. Theory Comput.} \textbf{2017}, \emph{13}, 4410--4420\relax
\mciteBstWouldAddEndPuncttrue
\mciteSetBstMidEndSepPunct{\mcitedefaultmidpunct}
{\mcitedefaultendpunct}{\mcitedefaultseppunct}\relax
\EndOfBibitem
\bibitem[Dreuw and Head-Gordon(2005)Dreuw, and Head-Gordon]{dreuw2005}
Dreuw,~A.; Head-Gordon,~M. Single-reference ab initio methods for the calculation of excited states of large molecules. \emph{Chem. Rev.} \textbf{2005}, \emph{105}, 4009--4037\relax
\mciteBstWouldAddEndPuncttrue
\mciteSetBstMidEndSepPunct{\mcitedefaultmidpunct}
{\mcitedefaultendpunct}{\mcitedefaultseppunct}\relax
\EndOfBibitem
\bibitem[Casida(1995)]{casida1995}
Casida,~M.~E. \emph{Recent Advances In Density Functional Methods: (Part I)}; World Scientific, 1995; pp 155--192\relax
\mciteBstWouldAddEndPuncttrue
\mciteSetBstMidEndSepPunct{\mcitedefaultmidpunct}
{\mcitedefaultendpunct}{\mcitedefaultseppunct}\relax
\EndOfBibitem
\bibitem[McLachlan and Ball(1964)McLachlan, and Ball]{mclachlan1964}
McLachlan,~A.~D.; Ball,~M.~A. Time-Dependent Hartree-Fock Theory for Molecules. \emph{Rev. Mod. Phys.} \textbf{1964}, \emph{36}, 844--855\relax
\mciteBstWouldAddEndPuncttrue
\mciteSetBstMidEndSepPunct{\mcitedefaultmidpunct}
{\mcitedefaultendpunct}{\mcitedefaultseppunct}\relax
\EndOfBibitem
\bibitem[Rowe(1968)]{rowe1968}
Rowe,~D. Equations-of-motion method and the extended shell model. \emph{Rev. Mod. Phys.} \textbf{1968}, \emph{40}, 153\relax
\mciteBstWouldAddEndPuncttrue
\mciteSetBstMidEndSepPunct{\mcitedefaultmidpunct}
{\mcitedefaultendpunct}{\mcitedefaultseppunct}\relax
\EndOfBibitem
\bibitem[Fabiano and Della~Sala(2012)Fabiano, and Della~Sala]{fabiano2012}
Fabiano,~E.; Della~Sala,~F. Accuracy of basis-set extrapolation schemes for DFT-RPA correlation energies in molecular calculations. \emph{Theor. Chem. Acc.} \textbf{2012}, \emph{131}, 1278\relax
\mciteBstWouldAddEndPuncttrue
\mciteSetBstMidEndSepPunct{\mcitedefaultmidpunct}
{\mcitedefaultendpunct}{\mcitedefaultseppunct}\relax
\EndOfBibitem
\bibitem[Ziegler \latin{et~al.}(2014)Ziegler, Krykunov, and Autschbach]{ziegler2014}
Ziegler,~T.; Krykunov,~M.; Autschbach,~J. Derivation of the RPA (random phase approximation) equation of ATDDFT (adiabatic time dependent density functional ground state response theory) from an excited state variational approach based on the ground state functional. \emph{J. Chem. Theory Comput.} \textbf{2014}, \emph{10}, 3980--3986\relax
\mciteBstWouldAddEndPuncttrue
\mciteSetBstMidEndSepPunct{\mcitedefaultmidpunct}
{\mcitedefaultendpunct}{\mcitedefaultseppunct}\relax
\EndOfBibitem
\bibitem[He{\ss}elmann(2015)]{hesselmann2015}
He{\ss}elmann,~A. Molecular excitation energies from time-dependent density functional theory employing random-phase approximation hessians with exact exchange. \emph{J. Chem. Theory Comput.} \textbf{2015}, \emph{11}, 1607--1620\relax
\mciteBstWouldAddEndPuncttrue
\mciteSetBstMidEndSepPunct{\mcitedefaultmidpunct}
{\mcitedefaultendpunct}{\mcitedefaultseppunct}\relax
\EndOfBibitem
\bibitem[{\v{C}}{\'\i}{\v{z}}ek and Paldus(1967){\v{C}}{\'\i}{\v{z}}ek, and Paldus]{cizek1967}
{\v{C}}{\'\i}{\v{z}}ek,~J.; Paldus,~J. Stability Conditions for the Solutions of the Hartree—Fock Equations for Atomic and Molecular Systems. Application to the Pi-Electron Model of Cyclic Polyenes. \emph{J. Chem. Phys.} \textbf{1967}, \emph{47}, 3976--3985\relax
\mciteBstWouldAddEndPuncttrue
\mciteSetBstMidEndSepPunct{\mcitedefaultmidpunct}
{\mcitedefaultendpunct}{\mcitedefaultseppunct}\relax
\EndOfBibitem
\bibitem[Thiel(2014)]{thiel2014}
Thiel,~W. Semiempirical quantum--chemical methods. \emph{WIREs Comput. Mol. Sci.} \textbf{2014}, \emph{4}, 145--157\relax
\mciteBstWouldAddEndPuncttrue
\mciteSetBstMidEndSepPunct{\mcitedefaultmidpunct}
{\mcitedefaultendpunct}{\mcitedefaultseppunct}\relax
\EndOfBibitem
\bibitem[Voityuk(2013)]{voityuk2013}
Voityuk,~A.~A. Intermediate neglect of differential overlap for spectroscopy. \emph{WIREs Comput. Mol. Sci.} \textbf{2013}, \emph{3}, 515--527\relax
\mciteBstWouldAddEndPuncttrue
\mciteSetBstMidEndSepPunct{\mcitedefaultmidpunct}
{\mcitedefaultendpunct}{\mcitedefaultseppunct}\relax
\EndOfBibitem
\bibitem[Ridley and Zerner(1976)Ridley, and Zerner]{ridley1976}
Ridley,~J.~E.; Zerner,~M.~C. Triplet states via intermediate neglect of differential overlap: benzene, pyridine and the diazines. \emph{Theor. Chim. Acta} \textbf{1976}, \emph{42}, 223--236\relax
\mciteBstWouldAddEndPuncttrue
\mciteSetBstMidEndSepPunct{\mcitedefaultmidpunct}
{\mcitedefaultendpunct}{\mcitedefaultseppunct}\relax
\EndOfBibitem
\bibitem[Bacon and Zerner(1979)Bacon, and Zerner]{bacon1979}
Bacon,~A.~D.; Zerner,~M.~C. An intermediate neglect of differential overlap theory for transition metal complexes: Fe, Co and Cu chlorides. \emph{Theor. Chim. Acta} \textbf{1979}, \emph{53}, 21--54\relax
\mciteBstWouldAddEndPuncttrue
\mciteSetBstMidEndSepPunct{\mcitedefaultmidpunct}
{\mcitedefaultendpunct}{\mcitedefaultseppunct}\relax
\EndOfBibitem
\bibitem[Zerner \latin{et~al.}(1980)Zerner, Loew, Kirchner, and Mueller-Westerhoff]{zerner1980}
Zerner,~M.~C.; Loew,~G.~H.; Kirchner,~R.~F.; Mueller-Westerhoff,~U.~T. An intermediate neglect of differential overlap technique for spectroscopy of transition-metal complexes. Ferrocene. \emph{J. Am. Chem. Soc.} \textbf{1980}, \emph{102}, 589--599\relax
\mciteBstWouldAddEndPuncttrue
\mciteSetBstMidEndSepPunct{\mcitedefaultmidpunct}
{\mcitedefaultendpunct}{\mcitedefaultseppunct}\relax
\EndOfBibitem
\bibitem[Mataga and Nishimoto(1957)Mataga, and Nishimoto]{mataga1957}
Mataga,~N.; Nishimoto,~K. Electronic structure and spectra of nitrogen heterocycles. \emph{Z. Phys. Chem.} \textbf{1957}, \emph{13}, 140--157\relax
\mciteBstWouldAddEndPuncttrue
\mciteSetBstMidEndSepPunct{\mcitedefaultmidpunct}
{\mcitedefaultendpunct}{\mcitedefaultseppunct}\relax
\EndOfBibitem
\bibitem[Cronstrand \latin{et~al.}(2000)Cronstrand, Christiansen, Norman, and {\AA}gren]{cronstrand2000}
Cronstrand,~P.; Christiansen,~O.; Norman,~P.; {\AA}gren,~H. Theoretical calculations of excited state absorption. \emph{Phys. Chem. Chem. Phys.} \textbf{2000}, \emph{2}, 5357--5363\relax
\mciteBstWouldAddEndPuncttrue
\mciteSetBstMidEndSepPunct{\mcitedefaultmidpunct}
{\mcitedefaultendpunct}{\mcitedefaultseppunct}\relax
\EndOfBibitem
\bibitem[Yeager \latin{et~al.}(1975)Yeager, Nascimento, and McKoy]{yeager1975}
Yeager,~D.~L.; Nascimento,~M. A.~C.; McKoy,~V. Some applications of excited-state-excited-state transition densities. \emph{Phys. Rev. A} \textbf{1975}, \emph{11}, 1168--1174\relax
\mciteBstWouldAddEndPuncttrue
\mciteSetBstMidEndSepPunct{\mcitedefaultmidpunct}
{\mcitedefaultendpunct}{\mcitedefaultseppunct}\relax
\EndOfBibitem
\bibitem[Klinkusch \latin{et~al.}(2009)Klinkusch, Saalfrank, and Klamroth]{klinkusch2009}
Klinkusch,~S.; Saalfrank,~P.; Klamroth,~T. Laser-induced electron dynamics including photoionization: A heuristic model within time-dependent configuration interaction theory. \emph{J. Chem. Phys.} \textbf{2009}, \emph{131}, 114304\relax
\mciteBstWouldAddEndPuncttrue
\mciteSetBstMidEndSepPunct{\mcitedefaultmidpunct}
{\mcitedefaultendpunct}{\mcitedefaultseppunct}\relax
\EndOfBibitem
\bibitem[Coccia \latin{et~al.}(2017)Coccia, Assaraf, Luppi, and Toulouse]{coccia2017}
Coccia,~E.; Assaraf,~R.; Luppi,~E.; Toulouse,~J. Ab initio lifetime correction to scattering states for time-dependent electronic-structure calculations with incomplete basis sets. \emph{J. Chem. Phys.} \textbf{2017}, \emph{147}, 014106\relax
\mciteBstWouldAddEndPuncttrue
\mciteSetBstMidEndSepPunct{\mcitedefaultmidpunct}
{\mcitedefaultendpunct}{\mcitedefaultseppunct}\relax
\EndOfBibitem
\bibitem[Frisch \latin{et~al.}(2016)Frisch, Trucks, Schlegel, Scuseria, Robb, Cheeseman, Scalmani, Barone, Petersson, Nakatsuji, Li, Caricato, Marenich, Bloino, Janesko, Gomperts, Mennucci, Hratchian, Ortiz, Izmaylov, Sonnenberg, Williams-Young, Ding, Lipparini, Egidi, Goings, Peng, Petrone, Henderson, Ranasinghe, Zakrzewski, Gao, Rega, Zheng, Liang, Hada, Ehara, Toyota, Fukuda, Hasegawa, Ishida, Nakajima, Honda, Kitao, Nakai, Vreven, Throssell, Montgomery, Peralta, Ogliaro, Bearpark, Heyd, Brothers, Kudin, Staroverov, Keith, Kobayashi, Normand, Raghavachari, Rendell, Burant, Iyengar, Tomasi, Cossi, Millam, Klene, Adamo, Cammi, Ochterski, Martin, Morokuma, Farkas, Foresman, and Fox]{gaussian}
Frisch,~M.~J.; Trucks,~G.~W.; Schlegel,~H.~B.; Scuseria,~G.~E.; Robb,~M.~A.; Cheeseman,~J.~R.; Scalmani,~G.; Barone,~V.; Petersson,~G.~A.; Nakatsuji,~H. \latin{et~al.}  Gaussian˜16 {R}evision {C}.01. 2016; Gaussian Inc. Wallingford CT\relax
\mciteBstWouldAddEndPuncttrue
\mciteSetBstMidEndSepPunct{\mcitedefaultmidpunct}
{\mcitedefaultendpunct}{\mcitedefaultseppunct}\relax
\EndOfBibitem
\bibitem[Becke(1993)]{becke1993}
Becke,~A.~D. {Density‐functional thermochemistry. III. The role of exact exchange}. \emph{J. Chem. Phys.} \textbf{1993}, \emph{98}, 5648--5652\relax
\mciteBstWouldAddEndPuncttrue
\mciteSetBstMidEndSepPunct{\mcitedefaultmidpunct}
{\mcitedefaultendpunct}{\mcitedefaultseppunct}\relax
\EndOfBibitem
\bibitem[Yanai \latin{et~al.}(2004)Yanai, Tew, and Handy]{yanai2004}
Yanai,~T.; Tew,~D.~P.; Handy,~N.~C. A new hybrid exchange--correlation functional using the Coulomb-attenuating method (CAM-B3LYP). \emph{Chem. Phys. Lett.} \textbf{2004}, \emph{393}, 51--57\relax
\mciteBstWouldAddEndPuncttrue
\mciteSetBstMidEndSepPunct{\mcitedefaultmidpunct}
{\mcitedefaultendpunct}{\mcitedefaultseppunct}\relax
\EndOfBibitem
\bibitem[Rostov \latin{et~al.}(2010)Rostov, Amos, Kobayashi, Scalmani, and Frisch]{rostov2010}
Rostov,~I.~V.; Amos,~R.~D.; Kobayashi,~R.; Scalmani,~G.; Frisch,~M.~J. Studies of the ground and excited-state surfaces of the retinal chromophore using CAM-B3LYP. \emph{J. Phys. Chem. B} \textbf{2010}, \emph{114}, 5547--5555\relax
\mciteBstWouldAddEndPuncttrue
\mciteSetBstMidEndSepPunct{\mcitedefaultmidpunct}
{\mcitedefaultendpunct}{\mcitedefaultseppunct}\relax
\EndOfBibitem
\bibitem[Sarkar \latin{et~al.}(2021)Sarkar, Boggio-Pasqua, Loos, and Jacquemin]{sarkar2021}
Sarkar,~R.; Boggio-Pasqua,~M.; Loos,~P.-F.; Jacquemin,~D. Benchmarking TD-DFT and wave function methods for oscillator strengths and excited-state dipole moments. \emph{J. Chem. Theory Comput.} \textbf{2021}, \emph{17}, 1117--1132\relax
\mciteBstWouldAddEndPuncttrue
\mciteSetBstMidEndSepPunct{\mcitedefaultmidpunct}
{\mcitedefaultendpunct}{\mcitedefaultseppunct}\relax
\EndOfBibitem
\bibitem[Sun \latin{et~al.}(2020)Sun, Zhang, Banerjee, Bao, Barbry, Blunt, Bogdanov, Booth, Chen, Cui, Eriksen, Gao, Guo, Hermann, Hermes, Koh, Koval, Lehtola, Li, Liu, Mardirossian, McClain, Motta, Mussard, Pham, Pulkin, Purwanto, Robinson, Ronca, Sayfutyarova, Scheurer, Schurkus, Smith, Sun, Sun, Upadhyay, Wagner, Wang, White, Whitfield, Williamson, Wouters, Yang, Yu, Zhu, Berkelbach, Sharma, Sokolov, and Chan]{pyscf}
Sun,~Q.; Zhang,~X.; Banerjee,~S.; Bao,~P.; Barbry,~M.; Blunt,~N.~S.; Bogdanov,~N.~A.; Booth,~G.~H.; Chen,~J.; Cui,~Z.-H. \latin{et~al.}  {Recent developments in the PySCF program package}. \emph{J. Chem. Phys.} \textbf{2020}, \emph{153}, 024109\relax
\mciteBstWouldAddEndPuncttrue
\mciteSetBstMidEndSepPunct{\mcitedefaultmidpunct}
{\mcitedefaultendpunct}{\mcitedefaultseppunct}\relax
\EndOfBibitem
\bibitem[Stewart(1990)]{stewart1990}
Stewart,~J.~J. MOPAC: a semiempirical molecular orbital program. \emph{J. Comput. Aided Mol. Des.} \textbf{1990}, \emph{4}, 1--103\relax
\mciteBstWouldAddEndPuncttrue
\mciteSetBstMidEndSepPunct{\mcitedefaultmidpunct}
{\mcitedefaultendpunct}{\mcitedefaultseppunct}\relax
\EndOfBibitem
\bibitem[Gieseking(2021)]{gieseking2021}
Gieseking,~R.~L. A new release of MOPAC incorporating the INDO/S semiempirical model with CI excited states. \emph{J. Comput. Chem.} \textbf{2021}, \emph{42}, 365--378\relax
\mciteBstWouldAddEndPuncttrue
\mciteSetBstMidEndSepPunct{\mcitedefaultmidpunct}
{\mcitedefaultendpunct}{\mcitedefaultseppunct}\relax
\EndOfBibitem
\bibitem[Moussa and Stewart()Moussa, and Stewart]{mopac}
Moussa,~J.~E.; Stewart,~J. J.~P. MOPAC. \url{https://http://openmopac.net}\relax
\mciteBstWouldAddEndPuncttrue
\mciteSetBstMidEndSepPunct{\mcitedefaultmidpunct}
{\mcitedefaultendpunct}{\mcitedefaultseppunct}\relax
\EndOfBibitem
\bibitem[Chi(1970)]{chi1970}
Chi,~B.~E. The eigenvalue problem for collective motion in the random phase approximation. \emph{Nucl. Phys. A} \textbf{1970}, \emph{146}, 449--456\relax
\mciteBstWouldAddEndPuncttrue
\mciteSetBstMidEndSepPunct{\mcitedefaultmidpunct}
{\mcitedefaultendpunct}{\mcitedefaultseppunct}\relax
\EndOfBibitem
\bibitem[Morassut \latin{et~al.}(2023)Morassut, Coccia, and Luppi]{morassut2023}
Morassut,~C.; Coccia,~E.; Luppi,~E. Quantitative performance analysis and comparison of optimal-continuum Gaussian basis sets for high-harmonic generation spectra. \emph{J. Chem. Phys.} \textbf{2023}, \emph{159}\relax
\mciteBstWouldAddEndPuncttrue
\mciteSetBstMidEndSepPunct{\mcitedefaultmidpunct}
{\mcitedefaultendpunct}{\mcitedefaultseppunct}\relax
\EndOfBibitem
\bibitem[Berkelbach(2018)]{berkelbach2018}
Berkelbach,~T.~C. Communication: Random-phase approximation excitation energies from approximate equation-of-motion coupled-cluster doubles. \emph{J. Chem. Phys.} \textbf{2018}, \emph{149}\relax
\mciteBstWouldAddEndPuncttrue
\mciteSetBstMidEndSepPunct{\mcitedefaultmidpunct}
{\mcitedefaultendpunct}{\mcitedefaultseppunct}\relax
\EndOfBibitem
\bibitem[Gensterblum \latin{et~al.}(1991)Gensterblum, Pireaux, Thiry, Caudano, Vigneron, Lambin, Lucas, and Kr\"atschmer]{gensterblum1991}
Gensterblum,~G.; Pireaux,~J.~J.; Thiry,~P.~A.; Caudano,~R.; Vigneron,~J.~P.; Lambin,~P.; Lucas,~A.~A.; Kr\"atschmer,~W. High-resolution electron-energy-loss spectroscopy of thin films of ${\mathrm{C}}_{60}$ on Si(100). \emph{Phys. Rev. Lett.} \textbf{1991}, \emph{67}, 2171--2174\relax
\mciteBstWouldAddEndPuncttrue
\mciteSetBstMidEndSepPunct{\mcitedefaultmidpunct}
{\mcitedefaultendpunct}{\mcitedefaultseppunct}\relax
\EndOfBibitem
\bibitem[Hertel \latin{et~al.}(1992)Hertel, Steger, de~Vries, Weisser, Menzel, Kamke, and Kamke]{hertel1992}
Hertel,~I.~V.; Steger,~H.; de~Vries,~J.; Weisser,~B.; Menzel,~C.; Kamke,~B.; Kamke,~W. Giant plasmon excitation in free ${\mathrm{C}}_{60}$ and ${\mathrm{C}}_{70}$ molecules studied by photoionization. \emph{Phys. Rev. Lett.} \textbf{1992}, \emph{68}, 784--787\relax
\mciteBstWouldAddEndPuncttrue
\mciteSetBstMidEndSepPunct{\mcitedefaultmidpunct}
{\mcitedefaultendpunct}{\mcitedefaultseppunct}\relax
\EndOfBibitem
\bibitem[Yoo \latin{et~al.}(1992)Yoo, Ruscic, and Berkowitz]{yoo1992}
Yoo,~R.; Ruscic,~B.; Berkowitz,~J. Vacuum ultraviolet photoionization mass spectrometric study of C6. \emph{J. Chem. Phys.} \textbf{1992}, \emph{96}, 911--918\relax
\mciteBstWouldAddEndPuncttrue
\mciteSetBstMidEndSepPunct{\mcitedefaultmidpunct}
{\mcitedefaultendpunct}{\mcitedefaultseppunct}\relax
\EndOfBibitem
\bibitem[Ju \latin{et~al.}(1993)Ju, Bulgac, and Keller]{ju1993}
Ju,~N.; Bulgac,~A.; Keller,~J.~W. Excitation of collective plasmon states in fullerenes. \emph{Phys. Rev. B} \textbf{1993}, \emph{48}, 9071--9079\relax
\mciteBstWouldAddEndPuncttrue
\mciteSetBstMidEndSepPunct{\mcitedefaultmidpunct}
{\mcitedefaultendpunct}{\mcitedefaultseppunct}\relax
\EndOfBibitem
\bibitem[Scully \latin{et~al.}(2005)Scully, Emmons, Gharaibeh, Phaneuf, Kilcoyne, Schlachter, Schippers, M\"uller, Chakraborty, Madjet, and Rost]{scully2005}
Scully,~S. W.~J.; Emmons,~E.~D.; Gharaibeh,~M.~F.; Phaneuf,~R.~A.; Kilcoyne,~A. L.~D.; Schlachter,~A.~S.; Schippers,~S.; M\"uller,~A.; Chakraborty,~H.~S.; Madjet,~M.~E. \latin{et~al.}  Photoexcitation of a Volume Plasmon in ${\mathrm{C}}_{60}$ Ions. \emph{Phys. Rev. Lett.} \textbf{2005}, \emph{94}, 065503\relax
\mciteBstWouldAddEndPuncttrue
\mciteSetBstMidEndSepPunct{\mcitedefaultmidpunct}
{\mcitedefaultendpunct}{\mcitedefaultseppunct}\relax
\EndOfBibitem
\end{mcitethebibliography}

\end{document}